\RequirePackage{fix-cm}
\documentclass[smallextended]{svjour3}       
\smartqed  
\usepackage{graphicx}
\usepackage{graphicx}
\usepackage{color, soul, xcolor}  
\usepackage{colortbl}
\usepackage{booktabs} 
\usepackage{enumitem}
\usepackage{caption}
\usepackage{amsmath}
\usepackage{listings}
\usepackage{balance}
\usepackage{caption}
\usepackage[caption=false,font=footnotesize,labelformat=simple]{subfig}
\usepackage[normalem]{ulem} 
\usepackage{framed}
\definecolor{shadecolor}{gray}{0.95}

\usepackage[linewidth=1pt]{mdframed} 
\usepackage[round]{natbib}   
\usepackage{supertabular} 
\usepackage{siunitx} 

\usepackage{pifont}
\newcommand{\starsymbol}{\ding{74}}%

\journalname{Empirical Software Engineering}
\begin{document}

\title{\textit{The `as Code' Activities}: Development Anti-patterns for Infrastructure as Code}
\titlerunning{Development Anti-patterns for Infrastructure as Code}  

\author{Akond Rahman  \and
        Effat Farhana \and 
        Laurie Williams 
}


\institute{Akond Rahman \at
              Department of Computer Science, Tennessee Technological University, Cookeville, TN, USA \\
              \email{arahman@tntech.edu}           
           \and
           Effat Farhana \at
              Department of Computer Science, North Carolina State University, Raleigh, NC, USA  \\
              \email{efarhan@ncsu.edu}
           \and
           Laurie Williams \at
              Department of Computer Science, North Carolina State University, Raleigh, NC, USA  \\
              \email{lawilli3@ncsu.edu}
}

\date{\color{red}{\textit{Pre-print accepted at the Empirical Software Engineering Journal} } }

\maketitle


\begin{abstract}

Context: The `as code' suffix in infrastructure as code (IaC) refers to applying software engineering activities, such as version control, to maintain IaC scripts. Without the application of these activities, defects that can have serious consequences may be introduced in IaC scripts. A systematic investigation of the development anti-patterns for IaC scripts can guide practitioners in identifying activities to avoid defects in IaC scripts. Development anti-patterns are recurring development activities that relate with defective IaC scripts. Goal: \textit{The goal of this paper is to help practitioners improve the quality of infrastructure as code (IaC) scripts by identifying development activities that relate with defective IaC scripts.} Methodology: We identify development anti-patterns by adopting a mixed-methods approach, where we apply quantitative analysis with 2,138 open source IaC scripts and conduct a survey with 51 practitioners. Findings: We observe five development activities to be related with defective IaC scripts from our quantitative analysis. We identify five development anti-patterns namely, `boss is not around', `many cooks spoil', `minors are spoiler', `silos', and `unfocused contribution'. Conclusion: Our identified development anti-patterns suggest the importance of `as code' activities in IaC because these activities are related to quality of IaC scripts.  


\end{abstract}

\section{Introduction}
\label{intro}

Infrastructure as code (IaC) is the practice of automatically defining and managing deployment environments, system configurations, and infrastructure through source code~\citep{Humble:2010:CD}. Before the inception of IaC, system operators used to create custom deployment scripts, which were not developed and maintained using a systematic software development process~\citep{pulling:strings:puppet:book}. The `as code' suffix refers to applying development activities considered to be good practices in software development, such as keeping scripts in version control, testing, and submitting code changes in small units~\citep{kief:iac:book}. With the availability of cloud computing resources such as Amazon Web Services~\footnote{https://aws.amazon.com/}, the development and maintenance of deployment scripts became complex, which motivated information technology (IT) organizations to treat their deployment scripts as regular software source code.  

IaC scripts are also referred to as configuration scripts~\citep{SharmaPuppet2016}~\citep{Humble:2010:CD} or configuration as code scripts~\citep{Rahman:RCOSE18}. IT organizations widely use commercial tools such as Puppet, to implement the practice of IaC~\citep{Humble:2010:CD}~\citep{JiangAdamsMSR2015}~\citep{ShambaughRehearsal2016}. These IaC tools provide programming syntax and libraries so that programmers can specify configuration and dependency information as scripts. For example, Puppet provides the `user resource' library~\citep{puppet-doc} to add, update, and delete users in local/remote servers and cloud instances. Instead of manually logging into servers and running commands, users can be configured on multiple servers by automatically running a single IaC script. The use of IaC scripts has resulted in benefits for IT organizations. For example, the use of IaC scripts helped the National Aeronautics and Space Administration (NASA) to reduce its multi-day patching process to 45 minutes~\citep{nasa:iac}. The Enterprise Strategy Group surveyed practitioners and reported the use of IaC scripts to help IT organizations gain 210\% in time savings and 97\% in cost savings on average~\citep{ESG:iac}. 


However, IaC scripts are susceptible to defects, similar to software source code~\citep{JiangAdamsMSR2015}. Defects in IaC scripts propagate at scale, causing serious consequences. For example, the execution of a defective IaC script erased home directories of $\mathtt{\sim}$270 users in cloud instances maintained by Wikimedia Commons~\citep{wiki:horror}. As another example, a defective IaC script resulted in an outage worth of 150 million USD for Amazon Web Services~\citep{aws:horror}. 

\begin{figure}[t]
\centering
\includegraphics[scale=0.8]{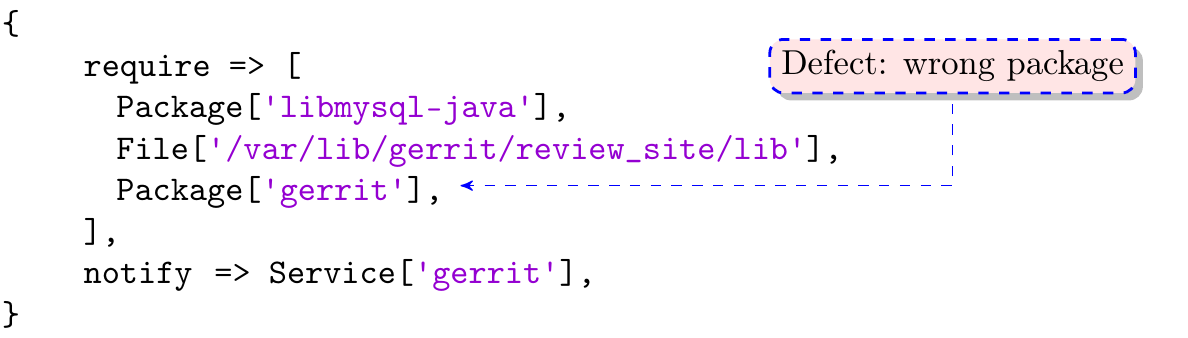}
\caption{Example of a defect included in the `825a073' commit, which is 460 lines long. }
\label{fig-intro}
\end{figure}

One strategy to prevent defects in IaC scripts is to identify development anti-patterns. A development anti-pattern is a recurring practice, which relates to a negative consequence in software development~\citep{ap:brown}. In our paper, a development anti-pattern is a recurring development activity that relates with defective scripts, as determined by empirical analysis. We can identify development activities that relate with defective IaC scripts by mining open source software (OSS) repositories~\citep{adams:release:nutshell}. Let us consider Figure~\ref{fig-intro}, which provides a code snippet from an actual commit (`825a073')~\footnote{https://github.com/Mirantis/puppet-manifests}. This 460-line long commit includes a defect. Reviewing large-sized commits, such as `825a073', is challenging~\citep{storey:codereview:2018}. As a result, defects existent in large commits may not be detected during inspection. This anecdotal example suggests a relationship that may exist between the activity of submitting large commits and defects that appear in IaC scripts. By mining metrics from OSS repositories, we can identify recurring development activities, such as submitting large-sized commits, and investigate their relationship with defective IaC scripts. 

In IaC development, submitting small sized commits is considered as an `as code' activity~\citep{kief:iac:book}. The activity of submitting small-sized commits is also applicable for software engineering in general~\citep{storey:codereview:2018}~\citep{Rigby:Apache:CodeReview}. Through systematic analysis, we can determine if non-adoption of the `as code' activities actually relate with defective IaC scripts. For example, if submitting large-sized commits is related with defective IaC scripts, then the identified relationship will underline the importance of the `as code' activity of submitting small sized commits. If we observe large-sized commits to recur across datasets, and observe large-sized commits to be related with defective IaC scripts, then we can identify submitting large-sized commits as a development anti-pattern. We quantify development activities using development activity metrics i.e. metrics that quantify an activity used to develop IaC scripts. For example, `commit size' is a development activity metric that quantifies how large or small are submitted commits are for IaC development.   

With development activity metrics we can quantify a set of activities used in IaC script development. We use each of these metrics as each one of them correspond to a development activity. Not all development activity metrics are actionable or can be mined from our repositories, and that is why we identify a specific set of development activity metrics. 

The metrics help us to quantitatively determine if a development activity is related with IaC script quality. But, to obtain relevance of our quantitative findings amongst practitioners we only cannot rely on quantitative findings. Hence, we survey practitioners involved in IaC script development. Survey results could reveal if our findings have relevance and also explain the reasons why practitioners may or may not find our findings to be relevant.     

Identification of development activity metrics can also be useful to prioritize inspection and testing efforts. We can construct prediction models with development activity metrics. Constructed prediction models can help practitioners automatically identify scripts that are likely to be defective. Instead of inspecting and testing all IaC scripts used by a team, practitioners from the team can prioritize their inspection and testing efforts for a subset of scripts. For general purpose programming languages defect prediction models are used to prioritize inspection and testing efforts in industry~\citep{dp:industry:menzies}~\citep{TURHAN:dp:industry}, and similar efforts can also be pursued for IaC scripts as well.     


\textit{The goal of this paper is to help practitioners improve the quality of infrastructure as code (IaC) scripts by identifying development activities that relate with defective IaC scripts.}

We answer the following research questions: 

\begin{itemize} 

\item{\textbf{RQ1}: How are development activity metrics quantitatively related with defective infrastructure as code scripts? }

\item{\textbf{RQ2}: What are practitioner perceptions on the relationship between development activity metrics and defective infrastructure as code scripts? }

\item{\textbf{RQ3}: How can we construct defect prediction models for IaC scripts using development activity metrics?}

\end{itemize}



We conduct quantitative analysis with 2,138 IaC scripts collected from 94 OSS repositories to investigate what development activity metrics relate with defective scripts. We hypothesize seven development activity metrics to show relationship with defective scripts. We consider these metrics for two reasons: (i) the metrics can provide actionable advice; and (ii) the metrics can be mined from OSS repositories. We investigate the relationship between the seven hypothesized metrics and defective IaC scripts using (i) statistical analysis, (ii) developer surveys and interviews, and (iii) prediction models. 

\textbf{Contributions}: We list our contributions as following: 

\begin{itemize}[leftmargin=*]
\item{A list of development anti-patterns for IaC scripts}; and 
\item{A set of defect prediction models constructed using development activity metrics}. 
\end{itemize}

We organize the rest of the paper as following: in Section~\ref{rel} we describe relevant related work. We describe our dataset construction process in Section~\ref{dataset}. We answer RQ1, RQ2, and RQ3 respectively, in Sections~\ref{rq1},~\ref{rq2}, and~\ref{rq3}. We discuss our findings in Section~\ref{discussion}. We discuss our limitations in Section~\ref{threats} and conclusions in Section~\ref{conclusion}.





\section{Related Work}
\label{rel}

Our paper is related to research studies that have focused on IaC technologies. Sharma et al.~\citep{SharmaPuppet2016} identified 13 code smells that may cause maintenance problems for Puppet scripts, such as missing default case and improper alignment. Jiang and Adams~\citep{JiangAdamsMSR2015} investigated the co-evolution of IaC scripts and other software artifacts and reported IaC scripts to experience frequent churn, similar to software source code. Weiss et al.~\citep{Weiss:Tortoise} proposed and evaluated `Tortoise', a tool that automatically corrects erroneous configurations in IaC scripts. Bent et al.~\citep{Bent:Saner2018:Puppet} proposed and validated nine metrics to detect maintainability issues in IaC scripts. Rahman et al.~\citep{Rahman:RCOSE18} investigated the questions that developers ask on Stack Overflow to identify the potential challenges developers face while working with Puppet. Rahman and Williams in separate studies characterized defective IaC scripts using text mining~\citep{me:icst2018:iac}, and by identifying source code properties~\citep{me:ist2019:code:properties}. In another work, Rahman et al.~\citep{me:icse2019:slic} identified 21,201 occurrences of security smells for IaC scripts that included 1,326 occurrences of hard-coded passwords. Rahman et al.~\citep{Rahman:2020:ACID} also constructed a defect taxonomy for IaC scripts that included eight defect categories.  

Our paper is related with prior research that have studied the relationships between development activity metrics and software quality. Meneely and Williams~\citep{Meneely:Linus} observed that development activity metrics, such as developer count, show relationship with vulnerable files. Rahman and Devanbu~\citep{Rahman:2013:ProcessBetter} observed that prediction models constructed using development activity metrics have higher prediction performance than models built with source code metrics. Pingzer et al.~\citep{Pinzger:Failure} observed that development activity metrics are related with software failures. Tufano et al.~\citep{penta:fix:commit} used development metrics, to predict fix-inducing code changes. 

The above-mentioned publications have not focused on empirical analysis that investigates what development anti-patterns may exist for IaC scripts. We address this research gap by taking inspiration from prior work that has used development activity metrics and apply it to the domain of IaC. IaC scripts use domain specific languages~\citep{ShambaughRehearsal2016}. Domain specific languages are different from general purpose programming languages with respect to syntax and semantics. The syntax and semantics of DSLs are different from GPLs~\citep{DSL:ENGG}~\citep{DSL:DIFF:HUDAK1998}~\citep{DSL:DIFF:VANWYK2007}. Furthermore, IaC scripts are often developed by system administrators who may not be well-versed on recommended software engineering practices, such as use of frequent commits. Differences in syntax, semantics, and development background necessitates to investigate to what extent existing development activity metrics are related with IaC script quality.


\section{Methodology Overview, Datasets, and Metrics}
\label{dataset}

In this section, first we provide an overview of methodology. Next, we discuss necessary datasets and metrics. 

\subsection{Methodology Overview}
\label{overview}

\begin{figure}[t]
\centering
\includegraphics[scale=0.75]{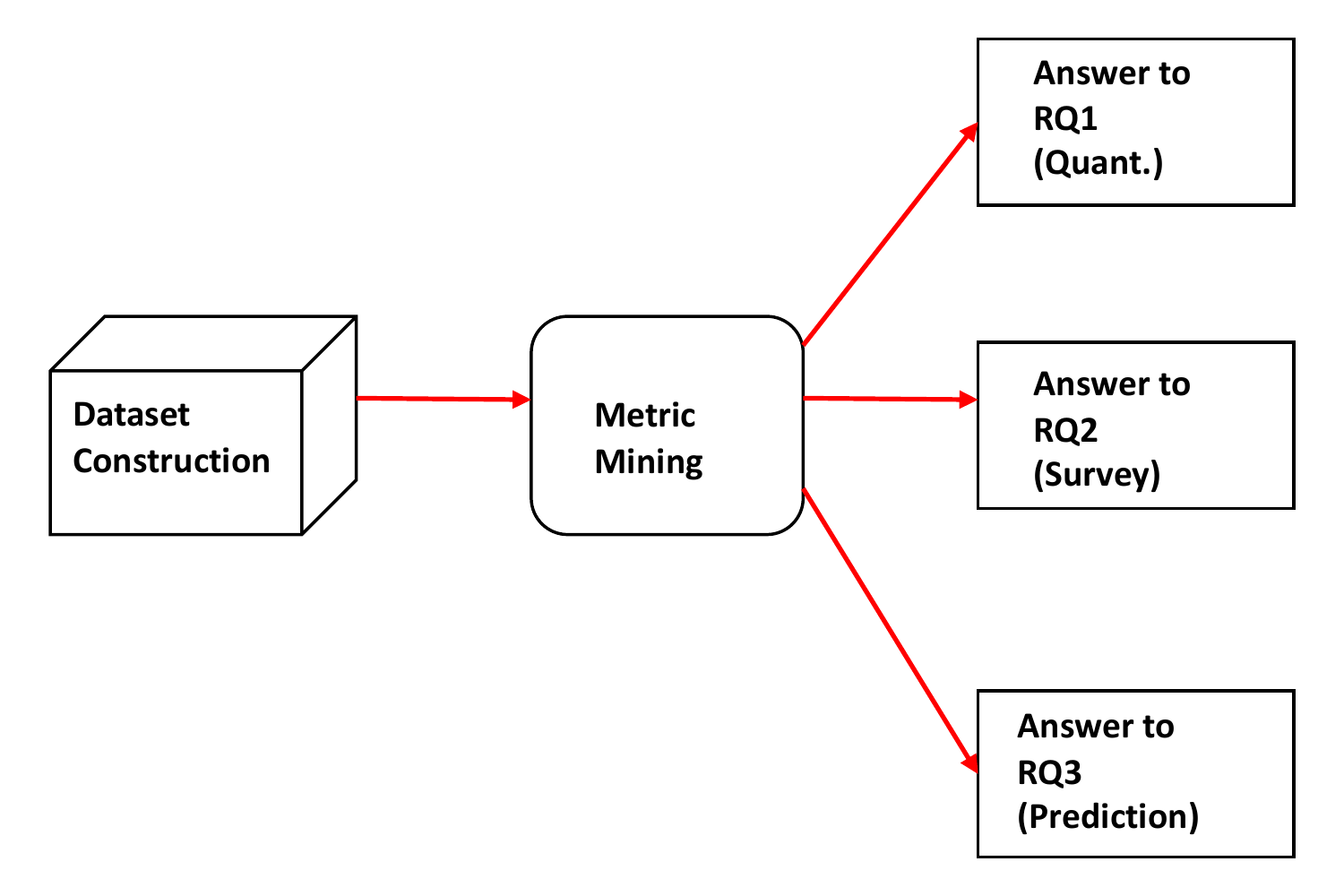}
\caption{A summary of our methodology to answer RQ1, RQ2, and RQ3. }
\label{fig-meth-summary}
\end{figure}

The practice of IaC includes the application of recommended development activities for traditional software engineering, such as submitting small-sized commits, in IaC script development. In our empirical study, we systematically investigate if development activity metrics that correlate with software quality as reported in prior work~\citep{Meneely:Linus}~\citep{Bird:Minor}, also show correlation with defective IaC scripts. We conduct our empirical study using a mixed-methods approach~\citep{storey:mixed:meth}, where we investigate the relationship between development activity metrics and defective scripts. \textit{First}, to answer \textbf{RQ1}, we derive development activity metrics using a scoping review~\citep{anderson2008:scoping}~\citep{scoping:original} i.e., a literature review in limited scope (Section~\ref{dataset-metrics}), and determine the relationship of the identified metrics using (i) Mann Whitney U Test~\citep{mann:whitney:original} and (ii) One way Multivariate Analysis of Variance (OMANOVA)~\citep{manova:book:huberty}. \textit{Second}, we answer \textbf{RQ2} to obtain practitioner perspective on our quantitative results from RQ1. We conduct an online survey and semi-structured interviews to synthesize practitioner perceptions. Our synthesis reveals the reasons why practitioners agree and disagree with our quantitative findings from RQ1. \textit{Third}, we answer \textbf{RQ3} by investigating if defect prediction models can be constructed using the development activity metrics which relate with defective scripts. The constructed prediction models can help practitioners to prioritize defective IaC scripts for further inspection and testing. A summary of our methodology is presented in Figure~\ref{fig-meth-summary}. 

To answer our research questions, we construct datasets in which there is a mapping between a defect and an IaC script. In this section, we describe the methodology to construct the datasets, and the metrics that we used to conduct our empirical study. First, we describe the methodology to construct our datasets. In our dataset, a defect is an imperfection that needs to be replaced or repaired, based upon the IEEE definition of defects~\citep{ieee:def}. A defect-related commit is a commit whose message indicates an action was taken to address a defect. We refer to an IaC script that is listed in a defect-related commit as a defective script. An IaC script that is listed in a commit, which is not defect-related is referred to as a neutral script. 




\subsection{Dataset Construction} 
\label{meth-ds}

We use Puppet scripts to construct our dataset because Puppet is considered one of the most popular tools to implement IaC~\citep{JiangAdamsMSR2015}~\citep{ShambaughRehearsal2016} and has been used by companies since 2005~\citep{propuppet:book}. We construct our datasets using OSS repositories collected from four organizations: Mirantis, Mozilla, Openstack, and Wikimedia Commons. We select repositories from these four organizations because these organizations create or use cloud-based infrastructure services, and our assumption is that an analysis of IaC scripts collected from these organizations could give us sufficient data to conduct our research study. Our assumption is that repositories collected from these four organizations will contain sufficient IaC scripts for analysis. We describe our steps to construct datasets below: 

\subsubsection{Repository Selection}
\label{repo-collect}

We select repositories needed for analysis by applying the following criteria:

\textbf{Criterion-1}: The repository must be available for download.

\textbf{Criterion-2}: At least 11\% of the files belonging to the repository must be IaC scripts. Jiang and Adams~\citep{JiangAdamsMSR2015} reported that in OSS repositories IaC scripts co-exist with other types of files, such as Makefiles and source code files. They observed a median of 11\% of the files to be IaC scripts. Our hypothesis is that we will be able to obtain repositories that contain sufficient amount of IaC scripts by using the cutoff of 11\%. 

\textbf{Criterion-3}: The repository must have at least two commits per month. Munaiah et al.~\citep{MunaiahCuration2017} used the threshold of at least two commits per month to determine which repositories have enough development activity for software organizations. We use this threshold to filter repositories with short development activity.

\subsubsection{Commit Message Processing}
\label{commit-collect}

For commit message processing we use: (i) commits that indicate modification of IaC scripts; and (ii) issue reports that are linked with the commits. We use commits because commits contain information on how and why a file was changed. We also use issue report summaries which can provide more insights on the issue. \textit{First}, we extract commits that were used to modify at least one IaC script. A commit lists the changes made on one or multiple files~\citep{maletic:commit:icpc2008}. \textit{Second}, we extract commit message i.e. the message of the commit identified from the previous step. The commit messages indicate why the changes were made to the corresponding files~\citep{maletic:commit:icpc2008}. \textit{Third}, if the commit message included a unique identifier that maps the commit to an issue in the issue tracking system, we extract the identifier and use that identifier to extract the summary of the issue. We use regular expression and the corresponding issue tracking API to extract the issue identifier. \textit{Fourth}, we combine the commit message with any existing issue summary to construct the message for analysis. We refer to the combined message as `extended commit message (ECM)' throughout the rest of the paper. We use the extracted ECMs to separate the defect-related commits from the non defect-related commits, as described in~\ref{defect-related-commit}.

\subsubsection{Determining Defect-related Commits}
\label{defect-related-commit}
We use defect-related commits to identify the defective IaC scripts. We apply qualitative analysis to determine which commits are defect-related commits using the following three steps:  

\textbf{Categorization Phase}: At least two raters with software engineering experience determine which of the collected commits are defect-related. We adopt this approach to mitigate the subjectivity introduced by a single rater. Each rater determine an ECM as defect-related if the ECM suggests a defect-related action is taken. For reference, we provide raters with an electronic handbook on IaC~\citep{puppet-doc} and the IEEE publication on anomaly classification~\citep{ieee:def}. The number of ECMs to which we observe agreements amongst the raters should be recorded and the Cohen's Kappa~\citep{cohens:kappa} score should be computed.

\textbf{Resolution Phase}: Raters can disagree if a commit is defect-related. To resolve such disagreements, we use an additional rater who we refer to as the `resolver'. Upon completion of this step, we can classify which commits and ECMs are defect-related. We determine a script to be defective if the script is modified in a defect-related commit.

\textbf{Member Checking}: To evaluate the ratings of the raters in the categorization and the resolution phase, we randomly select 50 ECMs for each dataset. We contact practitioners and ask if they agree to our categorization of ECMs. High agreement between the raters' categorization and practitioners' feedback is an indication of how well the raters performed. The percentage of ECMs to which practitioners agreed upon and the Cohen's Kappa score should be computed.


\subsection{Summary of Dataset Construction}
\label{dataset-results}

We apply the three selection criteria presented in Section~\ref{repo-collect} to identify repositories for our analysis. We describe how many of the repositories satisfied each of the three criterion in Table~\ref{table-criteria-dataset}. Each row corresponds to the count of repositories that satisfy each criterion. For example for Mirantis 26 repositories satisfy Criterion-1. Altogether, we obtain 94 repositories to extract Puppet scripts from. 

\begin{table}[]
\centering
\setlength\abovecaptionskip{-1pt}
\caption{Selection Criteria to Construct Defect Datasets}
\label{table-criteria-dataset}
{\footnotesize
\begin{tabular}{ p{1.75cm}  p{1.0cm} p{1.0cm} p{1.0cm} p{1.0cm} }
\hline
\textbf{Criteria}   & Mirantis & Mozilla & Openstack & Wikimedia \\
\hline
\textbf{Criterion-1} & 26 & 1,594 & 1,253 & 1,638 \\
\textbf{Criterion-2} & 20 & 2     & 61 & 11 \\
\textbf{Criterion-3} & 20 & 2     & 61 & 11 \\
\hline
\textbf{Final}      & 20 & 2     & 61 & 11 \\
\hline
\end{tabular}
}
\end{table}


We report summary statistics on the collected repositories in Table~\ref{table-defect-dataset}. For example, for Mirantis we collect 180 Puppet scripts that map to 1,021 commits. Of these 1,021 commits, 82 commits include identifiers for bug reports. We identify 33.7\% of the 1,021 commits as defect-related commits, and 53.3\% of the 180 scripts as defective.   

\begin{table*}[]
\centering
\caption{Statistics of Four Datasets}
\label{table-defect-dataset}
{\footnotesize
\begin{tabular}{ p{2.5cm} p{1.8cm} p{1.8cm} p{1.8cm} p{1.8cm} }
\hline
\textbf{Attributes}  & Mirantis & Mozilla & Openstack & Wikimedia \\
\hline
\textbf{Puppet Scripts} & 180 & 299  & 1,363 & 296 \\
\textbf{Commits with Puppet Scripts} & 1,021  & 3,074  & 7,808  & 972  \\
\textbf{Commits with Report IDs} & 82  of 1021, 8.0\% & 2764 of 3074, 89.9\% & 2252 of 7808, 28.8\% & 210 of 972, 21.6\% \\
\textbf{Defect-related Commits} &  344 of 1021, 33.7\% & 558 of 3074, 18.1\%  & 1987 of 7808, 25.4\% & 298 of 972, 30.6\% \\
\textbf{Defective Puppet Scripts} & 96 of 180, 53.3\% & 137 of 299, 45.8\%  & 793 of 1363, 58.2\% & 160 of 296, 54.0\% \\
\hline
\end{tabular}
}
\end{table*}

\begin{table}[]
\centering
\setlength\abovecaptionskip{-1pt}
\caption{Member Checking Phase}
\label{table-dataset-practitioner}
{\footnotesize
\begin{tabular}{ p{2cm} p{1.35cm} p{1.25cm} p{1.35cm} p{1.25cm} }
\hline
\textbf{Property}              & Mirantis & Mozilla & Openstack & Wikimedia \\
\hline
\textbf{Contacts}              & 5      & 6      & 10     & 7 \\
\textbf{Agreement}             & 91.0\% & 94.0\% & 92.0\% & 98.0\% \\
\textbf{Cohen's $\kappa$}      & 0.8 & 0.9 & 0.8 & 0.9 \\
\textbf{Interpretation}        & Substantial & Almost Perfect & Substantial & Almost Perfect \\
\hline
\end{tabular}
}
\end{table}

We categorize ECMs to classify which collected commits are defect-related, using the following three phases.  

\textbf{Categorization Phase}: We describe the categorization phase for the four datasets: 

\textit{Mirantis}: We recruit students in a graduate course related to software engineering titled `Software Security' via e-mail. The course was conducted in Fall 2017 at North Carolina State University. The number of students in the class was 58, and 32 students agreed to participate. The average experience of the 32 students in software engineering is two years. On average, each student took 2.1 hours to categorize the 200 ECMs. We randomly distribute the 1,021 ECMs amongst the students such that each ECM is rated by at least two students. 

\textit{Mozilla}: Two graduate students separately apply qualitative analysis on 3,074 ECMs. The first and second rater, respectively, have experience in software engineering of three and two years. The first and second rater, respectively, took 37.0 and 51.2 hours to complete the categorization.

\textit{Openstack}: Two graduate students, separately, apply qualitative analysis on 7,808 ECMs from Openstack repositories. The first and second rater, respectively, have software engineering experience of two and one years. The first and second rater completed the categorization of the 7,808 ECMs, respectively, in 80 and 130 hours.

\textit{Wikimedia}: We recruit students in a graduate course related to software engineering titled `Software Security' via e-mail. The course was conducted in Fall 2016 at North Carolina State University. The number of students in the class was 74, and 54 students agreed to participate. We randomly distribute the 972 ECMs amongst the students such that each ECM is rated by at least two students. The average experience of the 54 students in software engineering is 2.3 years. On average, each student took 2.1 hours to categorize the 140 ECMs.

\textbf{Resolution Phase}: The first author of the paper is the resolver who resolved disagreements for all four datasets. The Cohen's Kappa is 0.5, 0.6, 0.5, and 0.7, respectively, for Mirantis, Mozilla, Openstack, and Wikimedia. Based on Landis and Koch's interpretation~\citep{Landis:Koch:Kappa:Range}, we observe `Fair', `Moderate', `Fair' and `Substantial' agreement amongst raters for the datasets Mirantis, Mozilla, Openstack, and Wikimedia. 

\textbf{Member Checking}: Following our methodology in Section~\ref{defect-related-commit}, we report the agreement level between the raters' and the practitioners' categorization for 50 randomly-selected ECMs in Table~\ref{table-dataset-practitioner}. The `Contacts' row presents how many developers we contacted. All contacted practitioners responded. We report the agreement level and Cohen's Kappa score for 50 ECMs respectively, in `Agreement' and `Cohen's $\kappa$' rows. 

We observe that the agreement between ours and the practitioners' categorization varies from 0.8 to 0.9, which is higher than that of the agreement between the raters in the Categorization Phase. One possible explanation can be related to how the resolver resolved the disagreements. The first author of the paper has industry experience in writing IaC scripts, which may help to determine categorizations that are consistent with practitioners, potentially leading to higher agreement. Another possible explanation can be related to the sample provided to the practitioners: the provided sample, though randomly selected, may include commit messages whose categorization are relatively easy to agree upon.  

Finally upon applying qualitative analysis, we identify defect-related commits. These defect-related commits list the changed Puppet scripts which we use to identify the defective Puppet scripts. We present the count of defect-related commits and defective Puppet scripts in Table~\ref{table-defect-dataset}.

\textbf{\textit{Dataset}}: Constructed datasets are available online~\footnote{https://figshare.com/s/c88ece116f803c09beb6}.


\subsection{Metrics}
\label{dataset-metrics} 

We consider seven development activity metrics: developer count, disjointness in developer groups, highest contributor's code, minor contributors, normalized commit size, scatteredness, and unfocused contribution. We consider these metrics for two reasons: (i) the metrics can provide actionable advice; and (ii) the metrics can be mined from the 94 repositories collected in Chapter~\ref{dataset}. The criterion to determine metric actionability is ``\textit{a metric has actionability if it allows a software manager to make an empirically informed decision}''~\citep{Meneely:Metrics:Tosem2013}. 

We obtain these metrics by performing a scoping review~\citep{munn2018:scoping}~\citep{scoping:original}~\citep{anderson2008:scoping} of International Conference on Software Engineering (ICSE) to identify papers published from 2008 to 2018 that are related to defect prediction, and identify papers published in the Conference on Computer and Communications Security (CCS) that are related to vulnerability prediction. We select these two conferences as ICSE and CCS are considered the premier venues, respectively, for software engineering and computer security. We apply scoping review, which is a reduced form of systematic literature review. In our paper, we investigate what development activities might be mentioned in prior literature, and scoping reviews can provide guidance on what development activity metrics could be useful for defect and vulnerability prediction. The first author conducted the scoping review. Upon application of scoping review, we identify seven metrics, each of which are described below with appropriate references:  


\textbf{Developer count:} Similar to prior research~\citep{Meneely:Linus}, we hypothesize defective scripts will be modified by more developers compared to that of neutral scripts. Developer count is actionable because the metric can provide actionable advice for software managers on how many developers could be allocated to an IaC script. Furthermore, using this metric practitioners can perform additional review and testing on scripts that are modified by many developers.    

\textbf{Disjointness in developer groups}: Disjointness measures how separate one developer group is to another. A developer group is a set of developers who work on the same set of scripts~\citep{Meneely:Linus}. Disjointness in developer groups is actionable because the metric can provide actionable advice for software managers on whether developers should work disjointly or with collaboration. We hypothesize that defective IaC scripts will be modified by developer groups that exhibit more disjointness compared to that of neutral scripts. Similar to prior research~\citep{Meneely:Linus}, with Equation~\ref{equ:max:edge:betw} we use the maximum of edge betweenness metric ($MAX\_EDG\_BTW$) in the developer network to measure the amount of disjointness between developer groups. 

\begin{figure}[]
\centering
\includegraphics[scale=0.9]{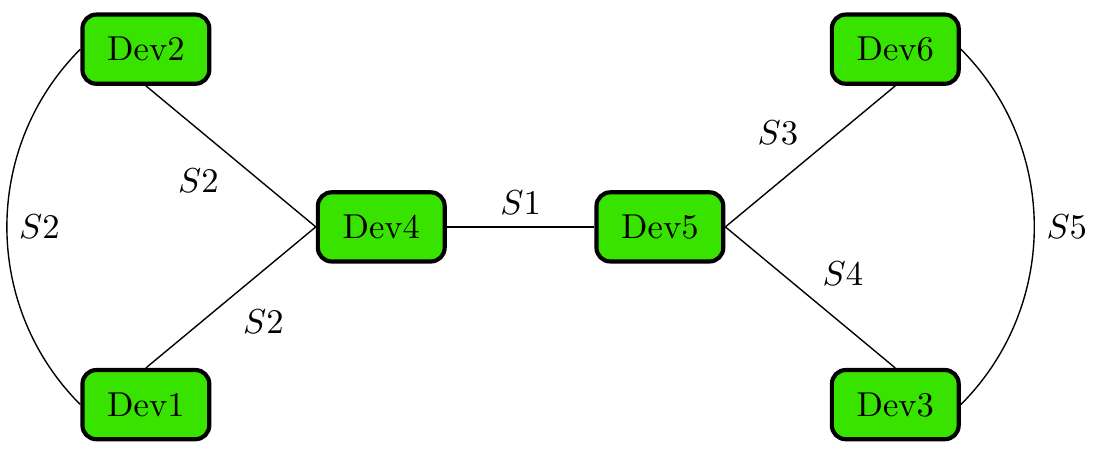}
\caption{A hypothetical example to demonstrate our calculation of disjointness between developers.}
\setlength\abovecaptionskip{-2pt}
\setlength\belowcaptionskip{-2pt}
\label{fig-meth-dev-net}
\end{figure}

We construct the graph with nodes, where each node is a developer. We only add an edge between two nodes if the same script is modified by two developers. Figure~\ref{fig-meth-dev-net} shows a developer network. As an example, in Figure~\ref{fig-meth-dev-net}, script $S1$ is modified by $Dev4$ and $Dev5$.  Our constructed developer network is an undirected, unweighted graph where each node corresponds to a developer. Each edge in the developer network connects two nodes if the same script is modified by developers that correspond to the two nodes of interest. In the developer network, a $MAX\_EDG\_BTW$ metric can tell empirically the groups of developers who are working together on the same script. We hypothesise that a defective script will have more developer groups hwo are disjoint compared to that of neutral scripts. For example, a script modified by two developer groups with a $MAX\_EDG\_BTW$ of 0.6 is likely to be more defective compared to that of a script, with $MAX\_EDG\_BTW$ of 0.2.

\begin{equation}
\tiny{MAX\_EDG\_BTW (e)}= max \sum\limits_{(a,b) \neq e} \frac{\text{\tiny{\# shortest paths between a and b that pass through e}}}{\text{\tiny{\# shortest paths between a and b}}}
\label{equ:max:edge:betw}
\end{equation}

\textbf{Highest contributor's code:} The highest contributor is the developer who authored the highest number of lines of code (LOC) for an IaC script. Highest contributor's code is actionable because the metric can provide actionable advice for software managers on whether or not contribution amount matters to the quality of IaC scripts. Similar to prior research~\citep{Devanbu:Process:Better}, we hypothesize that the highest contributor's code is significantly smaller in defective scripts compared to that of neutral scripts, which implies that developers who do not contribute majority of the scripts could introduce defects in IaC scripts. We calculate the contribution by the highest contributor ($HIGHEST\_CONTRIB\_CODE$) using Equation~\ref{equ:highest}.  

\begin{equation}
\footnotesize{HIGHEST\_CONTRIB\_CODE} = \frac{\text{\footnotesize{number of lines authored by the highest contributor}}}{\text{\footnotesize{lines of code}}}
\label{equ:highest}
\end{equation}

\textbf{Minor contributors:} Minor contributors is a subset of the developers who modify $\le5\%$ of total code for a script. Minor contributors is actionable because the metric can provide actionable advice for software managers on whether or not contribution amount matters to the quality of IaC scripts. Prior research~\citep{Devanbu:Process:Better} reported that scripts that are modified by minor contributors are more susceptible to defects. We hypothesize that the number of minor contributors are significantly larger for defective scripts compared to that of neutral scripts.   

\textbf{Normalized commit size}: Rahman and Devanbu~\citep{Devanbu:Process:Better} used normalized commit size to construct defect prediction models. Normalized commit size is actionable because the metric can provide actionable advice for software managers on whether IaC scripts should be developed by making small-sized commits or large-sized commits. We hypothesize that large-sized commits are more likely to appear in defective scripts than neutral scripts. We measure commit size using Equation~\ref{equ:change:size}, where we compute total lines added and deleted in all commits, and normalize by total commit count for a script.  

\begin{multline}
\footnotesize{Norm\_commit\_size} = 
\frac{\sum_{i=1}^{C}\text{\footnotesize{Total lines added/deleted in commit $i$}}}{\text{\footnotesize{Total number of commits}}}
\label{equ:change:size}
\end{multline}

\textbf{Scatteredness:} Scatteredness is actionable because the metric can provide actionable advice for software managers on whether or not submitted code changes should be spread out across a script or should be grouped together in a specific location of a script. Based upon findings from Hassan~\citep{Hassan:Entropy}, we hypothesize defective scripts will include more changes that are spread throughout the script when compared to neutral scripts. In Equation~\ref{equ:pi}, we calculate $x_i$, which we use in Equation~\ref{equ:scatter} to quantify scatteredness of a script. 

As an example, let us assume $Script\#1$ has 10 LOC and six commits. $Script\#2$ has seven LOC with four commits. For $Script\#1$, three modifications are made to line\#6 and 7 each. For $Script\#2$, line\#1, 2, 6, and 7 are modified once. According to Equation~\ref{equ:scatter}, the scatteredness score for $Script\#1$ and $Script\#2$ are respectively, 0.8 and 2.0.  

\begin{equation}
\footnotesize{x_i} = \frac{\text{\footnotesize{number of times line $i$ is modified}}}{\text{\footnotesize{number of commits in the script}}}
\label{equ:pi}
\end{equation}

\begin{equation}
\footnotesize{Scatteredness}= -\sum_{i=1}^{N}(x_ilog_2x_i)
\label{equ:scatter}
\end{equation}

\textbf{Unfocused contribution:} Unfocused contribution occurs when a script is changed by many developers who are also making many changes to other scripts~\citep{Pinzger:Failure}. Prior research has reported unfocused contribution to be related with software vulnerabilities~\citep{Meneely:Linus}. Unfocused contribution is actionable because the metric can provide actionable advice for software managers on whether or not multiple developers should modify multiple scripts. Unfocused contribution can also help practitioners identify scripts that require further inspection and testing. We hypothesize that defective scripts will be modified more through unfocused contributions compared to that of neutral scripts. 

\begin{figure}[]
\centering
\includegraphics[scale=0.9]{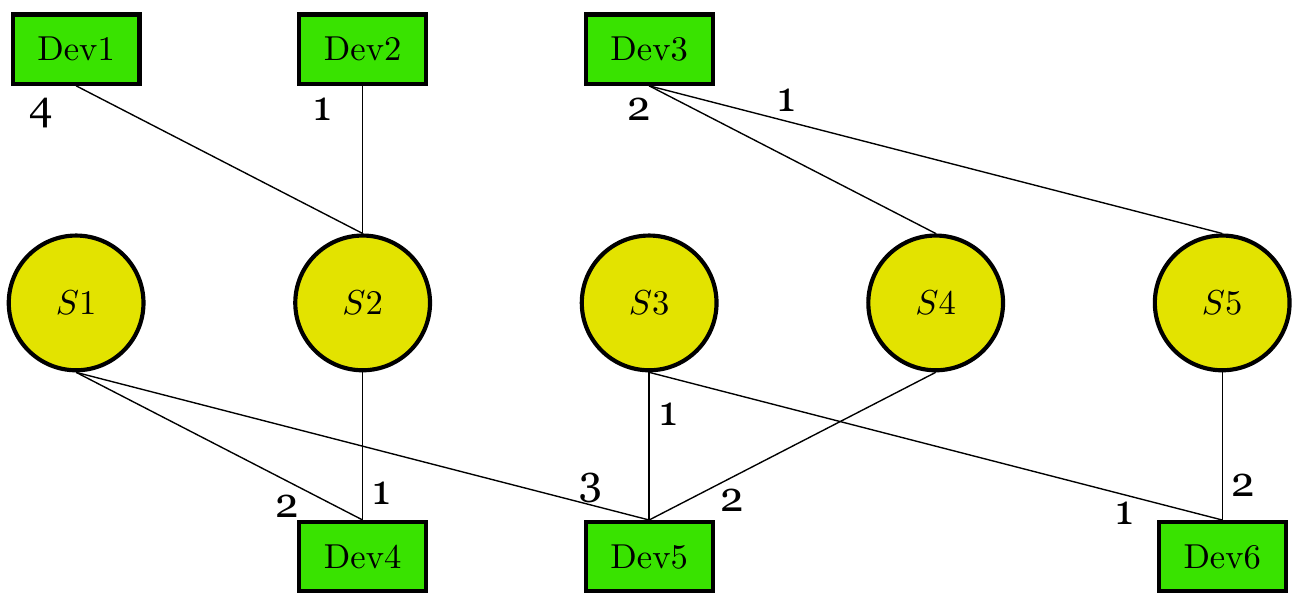}
\caption{A hypothetical example to demonstrate our calculation of unfocused contribution.}
\label{fig-meth-col-net}
\end{figure}

Similar to prior work~\citep{Meneely:Linus}, we use a graph called a contribution network, which is an un-directed, weighted graph. As shown in Figure~\ref{fig-meth-col-net}, two types of nodes exist in a contribution network: script (circle) and developer (rectangle). When a developer modifies a script, an edge between that developer and that script will exist. No edges are allowed between developers or between scripts. The number of commits the developer makes to a script is the weight for that edge. While determining unfocused contribution we use shortest paths that pass through a script. While determining shortest paths edge weights are accounted. 

We use the betweenness centrality metric ($BETW\_CENT$) of the contribution network to measure unfocused contribution using Equation~\ref{equ:betw:centr}. Equation~\ref{equ:betw:centr} presents the formula to compute betweenness centrality for script $x$, where $a$ and $b$ are developer nodes. A script with high betweenness indicates that the script has been modified by many developers who have made changes to other scripts. 

We use Figure~\ref{fig-meth-col-net} to provide an example for our betweenness centrality metric using S5. According to Figure~\ref{fig-meth-col-net}, $S5$ is modified by two developers, $Dev3$ and $Dev6$. There are two paths that connect $Dev3$ and $Dev6$: (i) $Dev3-S5-Dev6$, and (ii) $Dev3-S4-Dev5-S3-Dev6$ with a path length of respectively, 3 and 6. Only one shortest path exists between $Dev3$ and $Dev6$, which  is $Dev3-S5-Dev6$. Only one shortest path exists between $Dev3$ and $Dev6$ that also involves $S5$, which is $Dev3-S5-Dev6$. According to Equation~\ref{equ:betw:centr}, $BETW\_CENT$ for script $S5$ is 1/1 = 1.0.     

\begin{equation}
\text{\tiny{$BETW\_CENT$ (x)}}= \sum\limits_{a \neq x \neq b } \frac{\text{\tiny{no. of shortest paths between a and b that pass through x}}}{\text{\tiny{no. of shortest paths between a and b}}}
\label{equ:betw:centr}
\end{equation}


\section{RQ1: How are development activity metrics quantitatively related with defective infrastructure as code scripts?}
\label{rq1}

In Sections~\ref{meth-rq1} and~\ref{res-rq1} we respectively, provide the methodology and results to answer RQ1. 

\subsection{Methodology to Answer RQ1}
\label{meth-rq1}

We answer RQ1 by quantifying the relationship between each metric and defective IaC scripts. To determine the relationship we compare distributions instead of correlation analysis. Correlation analysis involves specifying cutoffs to specify if two features are strongly correlated or not~\cite{Tan:2005:IDM}. Application of correlation would require us, the authors, to decide if a development activity metric is strongly or weakly correlated with defects in IaC scripts. Using our judgement to determine correlations is susceptible to bias, and hence, we selected distribution comparison to determine the relationship between each metric and defective scripts. 

We use the seven development activity metrics identified from Section~\ref{dataset-metrics}. For each metric, we compare the distribution for that metric between defective and neutral scripts.

For metrics developer count, disjointness in developer groups, minor contributors, normalized commit size, scatteredness, and unfocused contribution, we state the following null and alternate hypothesis:  

\begin{itemize}
    \item{\textit{Null}:      the metric is not larger for defective scripts than neutral scripts.} 
    \item{\textit{Alternate}: the metric is larger for defective scripts than neutral scripts.} 
\end{itemize}

For highest contributor's code, we state the following null and alternate hypothesis:  

\begin{itemize}
    \item{\textit{Null}:      highest contributor's code is not larger for neutral scripts than defective scripts.} 
    \item{\textit{Alternate}: highest contributor's code is larger for neutral scripts than defective scripts.} 
\end{itemize}

We reject the null hypothesis for the seven metrics using two approaches: 
\begin{itemize}
    \item{Mann Whitney U Test: We use the Mann-Whitney U test~\citep{mann:whitney:original} to compare the metric values for defective and neutral scripts. The Mann-Whitney U test is non-parametric compares two distributions and states if one distribution is significantly larger or smaller than the other. } 
    
    \item{ One way multivariate analysis of variance (OMANOVA): One Way Multivariate analysis of variance (OMANOVA) is a statistical approach to compare multivariate means~\citep{manova:book:huberty}. OMANOVA is used to measure the impact of one or more independent variables on two or more dependent variables or outcomes. To measure the impact the OMANOVA test compares if the means for each independent variable is significantly different for the multiple outcomes. If the mean is significantly different, then we can conclude that a correlation exists between a certain independent variable and the outcome. The significance of difference in means is quantified using p-values. In OMANOVA analysis the impact of multiple independent variables can be tested simultaneously. While measuring the impact for one independent variable, the impact of other independent variables is accounted by computing how significantly different the mean values are.       
    
    We apply OMANOVA to sanity check the results from Mann Whitney U test as size and age of scripts can also impact our two possible outcomes: defective script and non-defective script. Using OMANOVA we can measure the impact for each of the seven metrics on our dependent variable. We can apply OMANOVA as it works for dependent variables with two or more outcomes.  
    
    OMANOVA expects independent variables to be normally distributed. If an independent variable is not normal then we apply $log_{e}(x+1)$ to transform the independent variables. We repeat the OMANOVA analysis for the seven metrics, where we also control for two additional independent variables: size and age of a script. We measure size and age of a script, respectively, in terms of lines of code and months between the first and last commit for the script. In our OMANOVA tests, if the mean for one of the seven metrics is significantly different even after including size and age, then we can conclude that our metric has a correlation with the outcomes even after considering the impact of size and age. 
    }
\end{itemize}

Following Cramer and Howitt's observations~\citep{cramer2004:pval:99}, for both Mann-Whitney U test and MANOVA we determine a metric to be significantly different if $p < 0.01$.  

We compute effect size using Cliff's Delta~\citep{cliff1993:original} for metrics that are significantly different between defective and non-defective scripts. Cliff's Delta is a non-parametric test~\citep{cliff1993:original} to compare the distribution of each metric between defective and neutral scripts. The Mann-Whitney U test shows if a relationship exists, but does not reveal the magnitude of differences~\citep{p:not:enough}. Effect size shows the magnitude of differences between two distributions. We use Romano et al.~\citep{Romano:CliffsCutoff2006}'s recommendations to interpret the observed Cliff's Delta values: the difference between two groups is `large' if Cliff's Delta is greater than 0.47. A Cliff's Delta value between 0.33 and 0.47 indicates a `medium' difference. A Cliff's Delta value between 0.14 and 0.33 indicates a `small' difference. Finally, a Cliff's Delta value less than 0.14 indicates a `negligible' difference.  

\subsection{Statistical Analysis Results}
\label{res-rq1} 

We organize the section in the following two subsections: first, we provide summary of metrics for all datasets in Section~\ref{res-rq1-metric-summary}. Next, we list the development activity metrics that relate with defective IaC scripts in Section~\ref{res-rq1-pval}.   

\subsubsection{Metric Summary for all Datasets}
\label{res-rq1-metric-summary}

We first provide minimum, maximum, and standard deviation of each metric for Mirantis, Mozilla, Openstack, and Wikimedia, respectively in Tables~\ref{table-rq1-dist-mir},~\ref{table-rq1-dist-moz},~\ref{table-rq1-dist-ost}, and~\ref{table-rq1-dist-wik}.

\subsubsection{Metrics that Relate with Defective Scripts}
\label{res-rq1-pval} 

In this section we provide the statistical test results for the seven development activity metrics. 

\textbf{Mann-Whitney U Test Results:} We use Mann-Whitney U Test to compare the distribution of each metric between defective and neutral scripts, and report the relationship between metrics and defective scripts. According to Table~\ref{table-res-rq1}, we observe a relationship to exist for five of the seven metrics: developer count, minor contributors, highest contributor's code, disjointness in developer groups, and unfocused contribution. The mean values of defective and neutral scripts are presented in the `(DE, NE)' column. The metrics for which $p < 0.01$, as determined by the Mann Whitney U test is indicated in grey cells. The metrics for which $p < 0.01$ for all datasets are followed by a star symbol (\starsymbol). We report the p-values in Table~\ref{table-res-rq1-pval}. 


\begin{table*}[]
\captionsetup{justification=centering}
\caption{Distribution of metrics for neutral and defective scripts in the Mirantis Dataset. Each tuple expresses the minimum, maximum, and standard deviation for each metric for both neutral and defective scripts.} 
\label{table-rq1-dist-mir}
\footnotesize
{
\begin{tabular}{p{4.5cm} p{2.5cm} p{2.5cm}  }
\hline
\textbf{Property}  & \textbf{Defective} & \textbf{Neutral}  \\
\hline
Developer count              & (1.0, 9.0, 1.2)    &  (1.0, 3.0, 0.6)    \\ 
\hline
Disjointness in dev. groups  & (0.0, 0.5, 0.1)    &  (0.0, 0.4, 0.2)           \\ 
\hline
Highest contrib. code        & (0.3, 1.0, 0.2)    &  (0.0, 1.0, 0.2)    \\ 
\hline
Minor contributors           & (0.0, 6.0, 0.8)    &  (0.0, 1.0, 0.2)       \\ 
\hline
Norm\_commit\_size           & (4.3, 191.7, 23.0) &  (0.0, 102.5, 21.0)   \\ 
\hline
Scatteredness                & (0.0, 5.9, 2.1)    &  (0.0, 4.8, 1.6)      \\ 
\hline
Unfocused contribution       & (1.0, 9.0, 1.1)    &  (0.0, 3.0, 0.5)     \\
\hline
\end{tabular}
}
\end{table*} 

\begin{table*}[]
\captionsetup{justification=centering}
\caption{Distribution of metrics for neutral and defective scripts in the Mozilla Dataset. Each tuple expresses the minimum, maximum, and standard deviation for each metric for both neutral and defective scripts.} 
\label{table-rq1-dist-moz}
\footnotesize
{
\begin{tabular}{p{4.5cm} p{2.5cm} p{2.5cm}  }
\hline
\textbf{Property}   & \textbf{Defective} & \textbf{Neutral} \\
\hline
Developer count              & (1.0, 25.0, 4.3)  &  (1.0, 8.0, 1.3)   \\ 
\hline
Disjointness in dev. groups  & (0.0, 0.6, 0.2)   &  (0.0, 0.4, 0.2)         \\ 
\hline
Highest contrib. code        & (0.2, 1.0, 0.2)   &  (0.0, 1.0, 0.2)   \\ 
\hline
Minor contributors           & (0.0, 12.0, 2.2)  &  (0.0, 6.0, 0.6)      \\ 
\hline
Norm\_commit\_size           &  (0.0, 65.0, 9.7) & (2.5, 62.3, 10.5)      \\ 
\hline
Scatteredness                &  (0.0, 4.4, 1.0)  & (0.2, 1.0, 0.2)         \\ 
\hline
Unfocused contribution       &  (0.0, 5.0, 1.0)  & (1.0, 8.0, 3.1)     \\
\hline
\end{tabular}
}
\end{table*} 

\begin{table*}[]
\captionsetup{justification=centering}
\caption{Distribution of metrics for neutral and defective scripts in the Openstack Dataset. Each tuple expresses the minimum, maximum, and standard deviation for each metric for both neutral and defective scripts.} 
\label{table-rq1-dist-ost}
\footnotesize
{
\begin{tabular}{p{4.5cm} p{2.5cm} p{2.5cm}  }
\hline
\textbf{Property}   & \textbf{Defective} & \textbf{Neutral} \\
\hline
Developer count              &  (1.0, 43.0, 4.0)   & (1, 11, 1.4)         \\ 
\hline
Disjointness in dev. groups  &  (0.0, 0.5, 0.1)    & (0.0, 0.5, 0.2)          \\ 
\hline
Highest contrib. code        &  (0.1, 1.0, 0.2)    & (0.2, 1.0, 0.2)       \\ 
\hline
Minor contributors           &  (0.0, 36.0, 3.1)   & (0.0, 7.0, 0.8)         \\ 
\hline
Norm\_commit\_size           &  (0.4, 277.0, 31.3) & (0.3, 207.0, 24.4)    \\ 
\hline
Scatteredness                &  (0.0, 6.7, 1.7)    & (0.0, 6.2, 1.2)          \\ 
\hline
Unfocused contribution       &  (1.0, 42.9, 4.0)   & (1, 11, 1.3)          \\
\hline
\end{tabular}
}
\end{table*} 

\begin{table*}[]
\captionsetup{justification=centering}
\caption{Distribution of metrics for neutral and defective scripts in the Wikimedia Dataset. Each tuple expresses the minimum, maximum, and standard deviation for each metric for both neutral and defective scripts.} 
\label{table-rq1-dist-wik}
\footnotesize
{
\begin{tabular}{p{4.5cm} p{2.5cm} p{2.5cm}  }
\hline
\textbf{Property}   & \textbf{Defective} & \textbf{Neutral} \\
\hline
Developer count              &  (1.0, 11.0, 1.7)  & (1.0, 5.0, 0.8)        \\ 
\hline
Disjointness in dev. groups  &  (0.0, 0.5, 0.2)   & (0.0, 0.4, 0.2)            \\ 
\hline
Highest contrib. code        &  (0.2, 1.0, 0.2)   & (0.4, 1.0, 0.1)       \\ 
\hline
Minor contributors           &  (0.0, 7.0, 1.1)   & (0.0, 2.0, 0.4)             \\ 
\hline
Norm\_commit\_size           &  (3.0, 170.5, 18.2)& (3.0, 135.5, 15.5)     \\ 
\hline
Scatteredness                &  (0.0, 5.9, 1.6)   & (0.0, 5.5, 1.4)        \\ 
\hline
Unfocused contribution       &  (1.0, 11.0, 1.5)  & (1.0, 5.0, 0.7)         \\
\hline
\end{tabular}
}
\end{table*} 




In Table~\ref{table-res-rq1-effect}, we present the Cliff's Delta values. Along with Cliff's Delta values we report Romano et al.~\citep{Romano:CliffsCutoff2006}'s interpretation of Cliff's Delta values: `N', `S', `M', and `L' respectively indicates `negligible', `small', `medium', and `large' difference. As indicated in bold, the metrics for which we observe `medium' or `large' differences between defective and neutral scripts across all datasets are: developer count and unfocused contribution. In short, relationship exists for five metrics, but the difference in metrics is large or medium for two metrics: developers and unfocused contribution, which suggests that for these two metrics the differences are more observable than the other three metrics. Practitioners can use the reported effect size measures as a strategy to prioritize which development anti-patterns they may act upon. For example, as the difference is `large' or `medium' for developer count, practitioners may thoroughly inspect IaC scripts modified by multiple developers.         

\begin{figure}
\centering 
\includegraphics[width=0.8\textwidth]{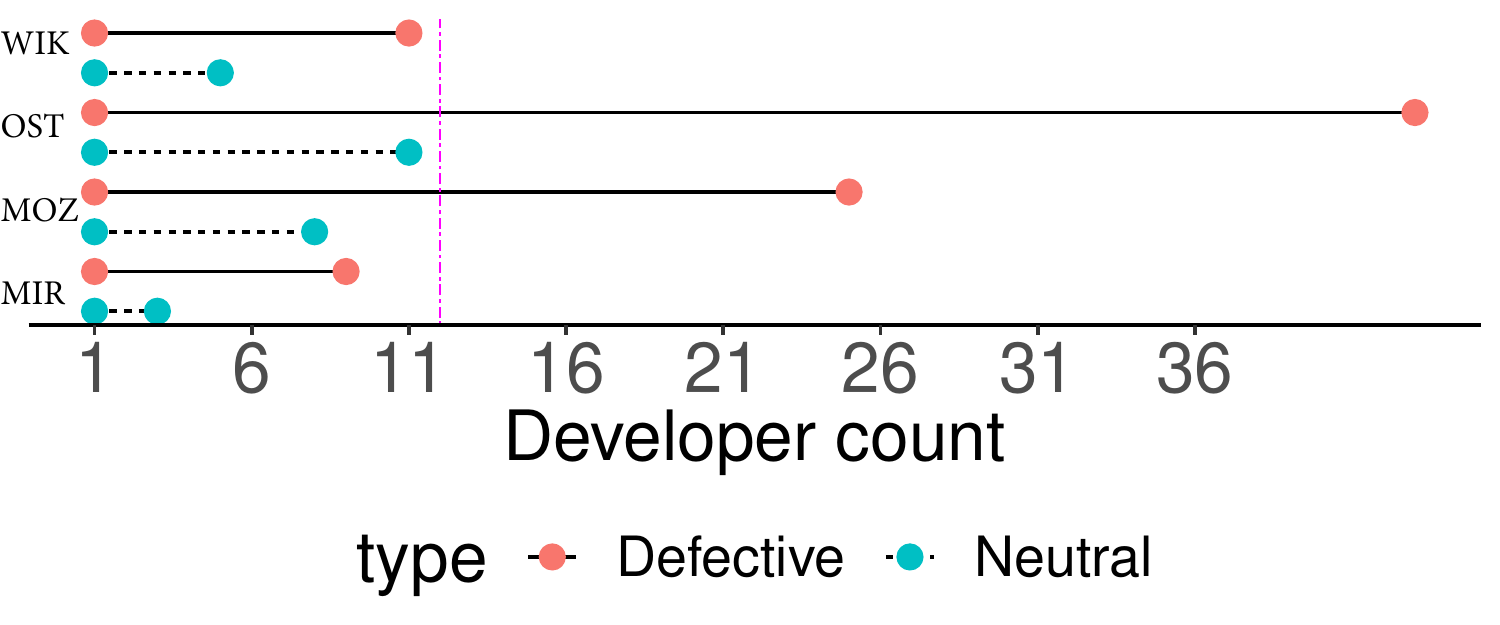}
\caption{Count of developers modifying scripts. Neutral scripts are modified by no more than 11 developers.}
\label{fig-res-devs-cutoff}
\end{figure}

From Table~\ref{table-res-rq1} we observe, on average, developer count is two times higher for defective scripts than neutral scripts. Figure~\ref{fig-res-devs-cutoff} presents the minimum and maximum number of developers who modify defective and neutral scripts. According to Figure~\ref{fig-res-devs-cutoff}, a neutral script may be modified by at most 11 developers. From Table~\ref{table-res-rq1} we also observe minor contributors to relate with defective scripts.

\begin{figure}
\centering 
\includegraphics[width=0.8\textwidth]{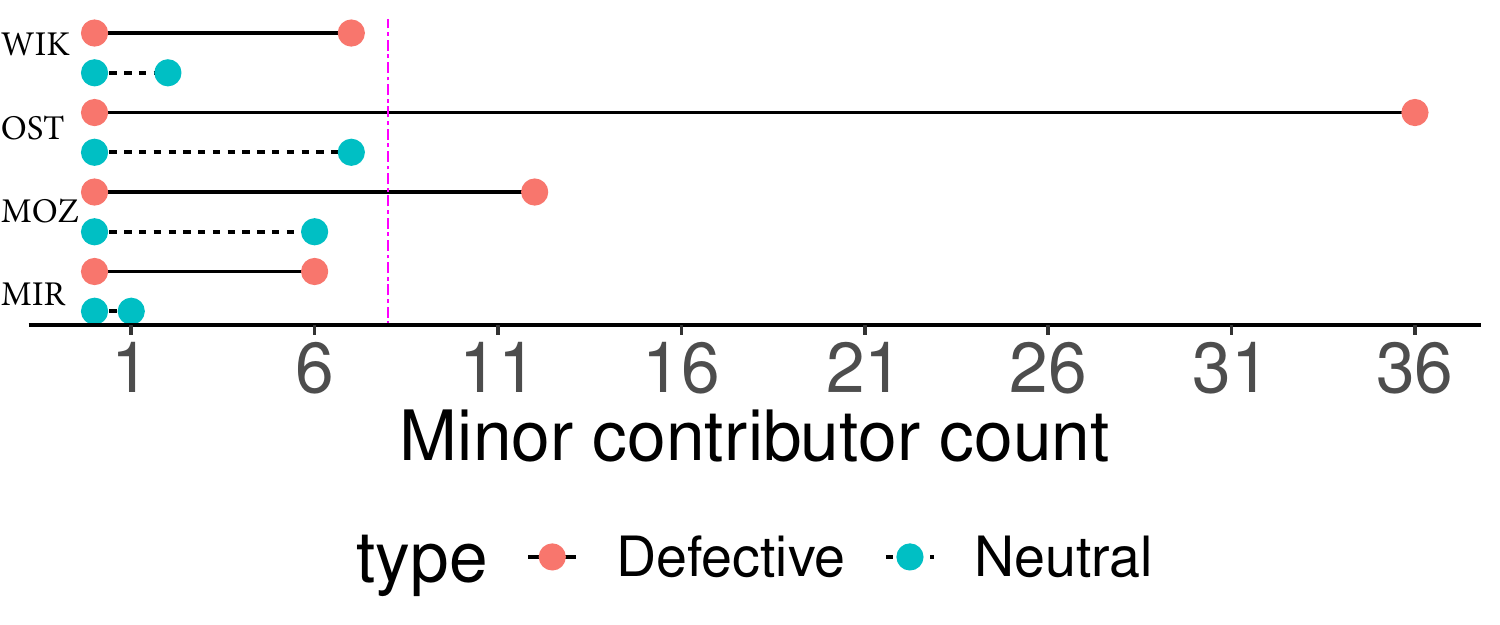}
\caption{No. of minor contributors modifying scripts. Neutral scripts are modified by no more than seven minor contributors.}
\label{fig-res-minors-cutoff}
\end{figure}

Figure~\ref{fig-res-minors-cutoff} presents the minimum and maximum number of minor contributors who modify defective and neutral scripts. We observe a neutral script may be modified by at most seven minor contributors. Also from Table~\ref{table-res-rq1-minmax}, we observe the minimum and maximum number of minor contributors for defective scripts are modified by $\ge 8$ minor contributors, and can be modified as many as 36 minor contributors. A complete breakdown of minimum and maximum values for developers and minor contributors is presented in Table~\ref{table-res-rq1-minmax}.  


\begin{table*}[]
\captionsetup{justification=centering}
\caption{Mean values for development activity metrics. The metrics for which $p < 0.01$ for all datasets are followed by a star symbol (\starsymbol).}
\label{table-res-rq1}  
{\footnotesize
\begin{tabular}{p{4.25cm} p{1.5cm}  p{1.3cm} p{1.3cm} p{1.3cm}  }
\hline
\textbf{Metric} & \textbf{MIR} & \textbf{MOZ} & \textbf{OST} & \textbf{WIK}\\
\hline 
 &  \textbf{DE, NE}  & \textbf{DE, NE}  & \textbf{DE, NE}  & \textbf{DE, NE}  \\
\hline
Developer count \starsymbol     & \cellcolor{lightgray} 2.5, 1.5  & \cellcolor{lightgray}4.1, 2.1  & \cellcolor{lightgray} 4.3, 2.2  & \cellcolor{lightgray} 2.6, 1.6  \\
\hline 
Disjointness in dev. groups \starsymbol & \cellcolor{lightgray} 0.4, 0.2   & \cellcolor{lightgray} 0.4, 0.3  & \cellcolor{lightgray} 0.4, 0.3  & \cellcolor{lightgray} 0.3, 0.2  \\
\hline 
Highest contrib. code \starsymbol        & \cellcolor{lightgray} 0.7, 0.8  & \cellcolor{lightgray}0.7, 0.8  & \cellcolor{lightgray} 0.6, 0.8  & \cellcolor{lightgray} 0.8, 0.9  \\
\hline 
Minor contributors \starsymbol   & \cellcolor{lightgray} 0.4, 0.0   & \cellcolor{lightgray}1.1, 0.2  & \cellcolor{lightgray} 1.6, 0.4  & \cellcolor{lightgray} 0.6, 0.1  \\
\hline 
Norm\_commit\_size  &  26.5, 24.1  & 14.0, 12.4  & 27.8, 28.3 & \cellcolor{lightgray} 18.8, 15.6  \\
\hline 
Scatteredness       & 2.6, 2.6   & \cellcolor{lightgray} 2.9, 2.2  & \cellcolor{lightgray} 3.4, 3.3  & \cellcolor{lightgray} 3.0, 2.4  \\
\hline 
Unfocused contribution \starsymbol & \cellcolor{lightgray} 2.5, 1.5   & \cellcolor{lightgray} 3.4, 1.8  & \cellcolor{lightgray} 4.3, 2.2  & \cellcolor{lightgray} 2.6, 1.6  \\
\hline 
\end{tabular}
}
\end{table*}  

\begin{table*}[]
\captionsetup{justification=centering}
\caption{Effect size values for development activity metrics. The metrics for which $p < 0.01$ for all datasets are followed by a star symbol (\starsymbol).} 
\label{table-res-rq1-effect}  
{\footnotesize
\begin{tabular}{p{4cm} p{1.25cm}  p{1.25cm} p{1.25cm}  p{1.25cm}  }
\hline
\textbf{Metric} & \textbf{MIR} & \textbf{MOZ} & \textbf{OST} & \textbf{WIK}\\
\hline
Developer count \starsymbol             & 0.55 (\textbf{L}) & 0.35 (\textbf{M})  & 0.40 (\textbf{M}) & 0.36 (\textbf{M})  \\
\hline 
Disjointness in dev. groups \starsymbol & 0.42 (\textbf{M})  & 0.21 (S)  & 0.21 (S) & 0.22 (S) \\
\hline 
Highest contrib. code \starsymbol       & 0.41 (\textbf{M})  & 0.24 (S)  & 0.37 (\textbf{M}) & 0.25 (S)  \\
\hline 
Minor contributors \starsymbol          & 0.26 (S) & 0.28 (S) & 0.30 (S) & 0.27 (S) \\
\hline 
Norm\_commit\_size                      & 0.12 (N) & 0.13 (N) & 0.06 (N) & 0.18 (S) \\
\hline 
Scatteredness                           & 0.06 (N) & 0.38 (\textbf{M}) & 0.15 (S) & 0.32 (S) \\
\hline 
Unfocused contribution \starsymbol      & 0.56 (\textbf{L}) & 0.35 (\textbf{M}) & 0.40 (\textbf{M}) &  0.36 (\textbf{M})  \\
\hline 
\end{tabular}
}
\end{table*}  


\begin{table*}[]
\captionsetup{justification=centering}
\caption{$p-value$ for development activity metrics. The metrics for which $p < 0.01$ for all datasets are followed by a star symbol (\starsymbol).} 
\label{table-res-rq1-pval}  
{\footnotesize
\begin{tabular}{p{4cm} p{1.25cm}  p{1.25cm} p{1.25cm}  p{1.25cm}  }
\hline
\textbf{Metric} & \textbf{MIR} & \textbf{MOZ} & \textbf{OST} & \textbf{WIK}\\
\hline
Developer count \starsymbol             & \num{2.1e-12} & \num{3.8e-08} & \num{1.5e-38} & \num{6.4e-09} \\
\hline 
Disjointness in dev. groups \starsymbol & \num{4.9e-10} & \num{5.6e-05} & \num{1.2e-21} & \num{2.3e-05} \\
\hline 
Highest contrib. code \starsymbol       & \num{8.7e-07} & 0.0001        & \num{2.3e-31} & \num{6.9e-05}\\
\hline 
Minor contributors \starsymbol          & \num{4.7e-06} & \num{2.4e-07} & \num{4.2e-26} & \num{3.9e-08} \\
\hline 
Norm\_commit\_size                      & 0.08          & 0.02          & 0.96          & 0.003 \\
\hline 
Scatteredness                           & 0.2           & \num{8.0e-09} & \num{7.2e-07} & \num{8.5e-07} \\
\hline 
Unfocused contribution \starsymbol      & \num{2.0e-11} & \num{1.6e-08} & \num{1.4e-37} & \num{6.3e-08}  \\
\hline 
\end{tabular}
}
\end{table*}  


\begin{table}[]
\captionsetup{justification=centering}
\caption{Minimum and maximum values for developer count and minor contributor count for defective and neutral IaC scripts}
\label{table-res-rq1-minmax}  
\begin{tabular}{p{2cm} | p{2cm} p{2cm}| p{2cm} p{2cm}}
\hline 
\textbf{Dataset} & \multicolumn{2}{c|}{\textbf{Developers}} & \multicolumn{2}{c}{\textbf{Minor contributors}}  \\
\hline 
            & Neutral & Defective  & Neutral & Defective \\
\hline 
Mirantis    & (1, 3)  & (1, 9)     & (0, 6)  & (0, 1) \\
Mozilla     & (1, 8)  & (1, 25)    & (0, 6)  & (0, 12) \\
Openstack   & (1, 11) & (1, 43)    & (0, 7)  & (0, 36) \\
Wikimedia   & (1, 5)  & (1, 11)    & (0, 2)  & (0, 7) \\
\hline 
\end{tabular}
\end{table}


\textbf{OMANOVA Results:} We provide results of our OMANOVA analysis in Tables~\ref{table-res-rebuttal-manova-devcount},~\ref{table-res-rebuttal-manova-disjoint},~\ref{table-res-rebuttal-manova-highest},~\ref{table-res-rebuttal-manova-minors},~\ref{table-res-rebuttal-manova-commit},~\ref{table-res-rebuttal-manova-scatter}, and~\ref{table-res-rebuttal-manova-unfocus}. In each of these tables we report the p-values, which show if the mean values are significantly different between defective and non-defective scripts. For example, in Table~\ref{table-res-rebuttal-manova-devcount} p-value is $< 0.01$ for size, age, and developer count, indicating size and age to correlate with defective scripts along with developer count.    

For all four datasets, the development activity metrics that correlate with defective scripts even when the effect of size and age is considered are: developer count, disjointness in developer groups, highest contributor code, minor contributors, and unfocused contributions. Our OMANOVA analysis is consistent with our Mann Whitney U Test results. 


\begin{table*}[]
\captionsetup{justification=centering}
\caption{OMANOVA results for all datasets when effect of size and age is considered with developer count. Developer count is significantly larger for defective scripts compared to neutral scripts for all datasets even when the effect of size and age is considered. Each cell represents a p-value for the metric. } 
\label{table-res-rebuttal-manova-devcount}  
{\footnotesize
\begin{tabular}{p{3.5cm} p{1.5cm}  p{1.5cm} p{1.5cm}  p{1.5cm}  }
\hline
\textbf{Metric} & \textbf{MIR}  & \textbf{MOZ}     & \textbf{OST}      & \textbf{WIK}\\
\hline
Size            & \num{3.4e-08} & $<$\num{2.2e-16} & \num{9.5e-11}     & \num{8.5e-15} \\
\hline 
Age             & \num{3.2e-13} & \num{3.5e-07}    &  $<$\num{2.2e-16} & \num{7.3e-07}\\
\hline 
Developer count & \num{9.3e-13} & \num{1.6e-09}    &  $<$\num{2.2e-16} & \num{2.2e-09} \\
\hline 
\end{tabular}
}
\end{table*}  


\begin{table*}[]
\captionsetup{justification=centering}
\caption{OMANOVA results for all datasets when effect of size and age is considered with disjointness in developer groups. Disjointness in developer groups is significantly larger for defective scripts compared to neutral scripts for all datasets even when the effect of size and age is considered. Each cell represents a p-value for the metric. } 
\label{table-res-rebuttal-manova-disjoint}  
{\footnotesize
\begin{tabular}{p{3.5cm} p{1.5cm}  p{1.5cm} p{1.5cm}  p{1.5cm}  }
\hline
\textbf{Metric}             & \textbf{MIR}  & \textbf{MOZ}     & \textbf{OST}      & \textbf{WIK}\\
\hline
Size                        & \num{3.4e-08} & $<$\num{2.2e-16} & \num{9.5e-11}     & \num{8.5e-15} \\
\hline 
Age                         & \num{3.2e-13} & \num{3.5e-07}    &  $<$\num{2.2e-16} & \num{7.3e-07}\\
\hline 
Disjointness in dev. groups & \num{1.1e-10} & \num{9.4e-05}    &  $<$\num{2.2e-16} & \num{3.6e-05} \\
\hline 
\end{tabular}
}
\end{table*}  


\begin{table*}[]
\captionsetup{justification=centering}
\caption{OMANOVA results for all datasets when effect of size and age is considered with highest contributor's code. Highest contributor's code is significantly larger for neutral scripts compared to defective scripts for all datasets even when the effect of size and age is considered. Each cell represents a p-value for the metric. } 
\label{table-res-rebuttal-manova-highest}  
{\footnotesize
\begin{tabular}{p{3.5cm} p{1.25cm}  p{1.5cm} p{1.5cm}  p{1.25cm}  }
\hline
\textbf{Metric}       & \textbf{MIR}  & \textbf{MOZ}     & \textbf{OST}      & \textbf{WIK}\\
\hline
Size                  & \num{3.4e-08} & $<$\num{2.2e-16} & \num{9.5e-11}     & \num{8.5e-15} \\
\hline 
Age                   & \num{3.3e-13} & \num{3.5e-07}    &  $<$\num{2.2e-16} & \num{7.3e-07}\\
\hline 
Highest contrib. code & 0.008         & 0.004            &  $<$\num{2.2e-16} & 0.0007 \\
\hline 
\end{tabular}
}
\end{table*}


\begin{table*}[]
\captionsetup{justification=centering}
\caption{OMANOVA results for all datasets when effect of size and age is considered with minor contributors. Minor contributors is significantly larger for defective scripts compared to neutral scripts for all datasets even when the effect of size and age is considered. Each cell represents a p-value for the metric. } 
\label{table-res-rebuttal-manova-minors}  
{\footnotesize
\begin{tabular}{p{3.5cm} p{1.25cm}  p{1.5cm} p{1.5cm}  p{1.25cm}  }
\hline
\textbf{Metric}    & \textbf{MIR}  & \textbf{MOZ}     & \textbf{OST}      & \textbf{WIK}\\
\hline
Size               & \num{3.4e-08} & $<$\num{2.2e-16} & \num{9.4e-11}     & \num{8.6e-15} \\
\hline 
Age                & \num{3.3e-13} & \num{3.5e-07}    &  $<$\num{2.2e-16} & \num{7.3e-07}\\
\hline 
Minor contributors & \num{9.2e-06} & \num{4.6e-08}    &  $<$\num{2.2e-16} & \num{4.2e-08} \\
\hline 
\end{tabular}
}
\end{table*}  


\begin{table*}[]
\captionsetup{justification=centering}
\caption{OMANOVA results for all datasets when effect of size and age is considered with normalized commit size. Normalized commit size is significantly larger for defective scripts compared to neutral scripts for none of the datasets when the effect of size and age is considered. Each cell represents a p-value for the metric. } 
\label{table-res-rebuttal-manova-commit}  
{\footnotesize
\begin{tabular}{p{3.5cm} p{1.25cm}  p{1.5cm} p{1.5cm}  p{1.25cm}  }
\hline
\textbf{Metric}        & \textbf{MIR}  & \textbf{MOZ}     & \textbf{OST}      & \textbf{WIK}\\
\hline
Size                   & \num{3.4e-08} & $<$\num{2.2e-16} & \num{9.5e-11}     & \num{8.5e-15} \\
\hline 
Age                    & \num{3.2e-13} & \num{3.5e-07}    &  $<$\num{2.2e-16} & \num{7.3e-07}\\
\hline 
Norm\_commit\_size     & 0.03          & 0.06             &  0.26             & 0.01 \\
\hline 
\end{tabular}
}
\end{table*}  


\begin{table*}[]
\captionsetup{justification=centering}
\caption{OMANOVA results for all datasets when effect of size and age is considered with scatteredness. Scatteredness is significantly larger for defective scripts compared to neutral scripts for one of the four datasets when the effect of size and age is considered. Each cell represents a p-value for the metric. } 
\label{table-res-rebuttal-manova-scatter}  
{\footnotesize
\begin{tabular}{p{3.5cm} p{1.25cm}  p{1.5cm} p{1.5cm}  p{1.25cm}  }
\hline
\textbf{Metric}        & \textbf{MIR}  & \textbf{MOZ}     & \textbf{OST}      & \textbf{WIK}\\
\hline
Size                   & \num{3.4e-08} & $<$\num{2.2e-16} & \num{9.5e-11}     & \num{8.5e-15} \\
\hline 
Age                    & \num{3.2e-13} & \num{3.5e-07}    &  $<$\num{2.2e-16} & \num{7.3e-07}\\
\hline 
Scatteredness          & 0.25          & 0.0001           &  0.04             & 0.02 \\
\hline 
\end{tabular}
}
\end{table*}  


\begin{table*}[]
\captionsetup{justification=centering}
\caption{OMANOVA results for all datasets when effect of size and age is considered with unfocused contribution. Unfocused contribution is significantly larger for defective scripts compared to neutral scripts for all datasets even when the effect of size and age is considered. Each cell represents a p-value for the metric. } 
\label{table-res-rebuttal-manova-unfocus}  
{\footnotesize
\begin{tabular}{p{3.5cm} p{1.25cm}  p{1.5cm} p{1.5cm}  p{1.25cm}  }
\hline
\textbf{Metric}        & \textbf{MIR}  & \textbf{MOZ}     & \textbf{OST}      & \textbf{WIK}\\
\hline
Size                   & \num{3.4e-08} & $<$\num{2.2e-16} & \num{9.5e-11}     & \num{8.5e-15} \\
\hline 
Age                    & \num{3.2e-13} & \num{3.5e-07}    &  $<$\num{2.2e-16} & \num{7.3e-07}\\
\hline 
Unfocused contribution & \num{9.3e-13} & \num{4.3e-05}    &  $<$\num{2.2e-16} & \num{2.3e-09} \\
\hline 
\end{tabular}
}
\end{table*}  

\begin{mdframed}
\textbf{Answer to RQ1}: Based on quantitative analysis, metrics related with defective scripts are: developer count, disjointness in developer groups, highest contributor's code, minor contributors, and unfocused contribution.       
\end{mdframed}

\section{RQ2: What are practitioner perceptions on the relationship between development activity metrics and defective infrastructure as code scripts?} 
\label{rq2}

In Sections~\ref{meth-rq2} and~\ref{res-rq2} we respectively report the methodology and results for RQ2

\subsection{Methodology}
\label{meth-rq2}

We answer RQ2 by using the metrics listed in Section~\ref{dataset-metrics}. We perform two tasks: (i) deploy a survey to practitioners; and (ii) conduct semi-structured interviews. In the survey, we asked the practitioners to what extent they agreed with a relationship between each of the seven development activity metric and defective IaC scripts. We constructed the survey following Kitchenham and Pfleeger's guidelines~\citep{kitchenham2008:survey} as follows: (i) use Likert scale to measure agreement levels: strongly disagree, disagree, neutral, agree, and strongly agree; (ii) add explanations related to the purpose of the study, how to conduct the survey, preservation of confidentiality, relevance to participants, and an estimate of the time to complete the questionnaire; (iii) administer a pilot survey to get initial feedback on the survey. Following Smith et al.~\citep{smith:chase2013:survey}'s  observations on software engineering survey incentives, we offer a drawing of five Amazon gift cards as an incentive for participation.

From the pilot study feedback with five graduate students, we add an open-ended question to the survey, where we ask participants to provide any other development activity that is related with quality but not represented in our set of seven metrics. We deploy the survey to 250 practitioners from July 12, 2018 to Jan 15, 2019. We collect developer e-mail addresses from the 94 repositories. Following IRB\#12598 protocol~\footnote{https://research.ncsu.edu/sparcs/compliance/irb/using-the-eirb-system/}, we distribute the survey to practitioners via e-mail. 


\begin{figure*}
\centering 
\includegraphics[width=0.95\textwidth]{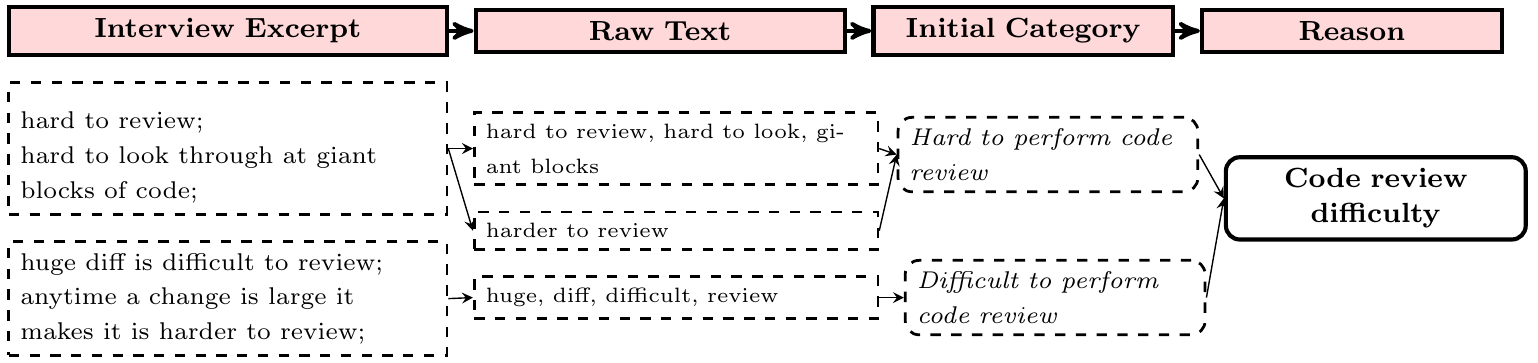}
\caption{Example of how we use qualitative analysis to determine reasons that attribute to practitioner perception.}
\label{fig-meth-rq2-gt}
\end{figure*}



\textbf{Semi-structured interview}: The results from the survey indicates the level of agreement from practitioners, but not the reasons that attribute to that perception. We conducted semi-structured interviews, i.e. an interview where practitioners are asked open-ended and closed questions to identify these reasons. We conduct all interviews over Skype or Google Hangouts. In each interview, we asked the interviewee to what extent they agree on the relationship between each metric and defective script using the Likert scale: strongly disagree, disagree, neutral, agree and strongly agree. Then, we asked about reasons that attribute to their perception. In the end, we asked interviewees if they would like to provide additional information about development activities related to defective scripts, but not included in the survey. We recruited the interviewees from the set of 250 practitioners to which we deployed the survey. We followed the guidelines provided by Hove and Anda~\citep{hove:anda:semi:interview} while conducting the interviews. We executed pilot interviews with a voluntary graduate student. During the interviews, we explained the purpose of the interview and explicitly mention that the interviewee's confidentiality will be respected.

\textbf{Qualitative analysis}: We transcribe the audio of each interview into textual content. From the textual content of semi-structured interviews, we apply a qualitative analysis technique called descriptive coding~\citep{saldana2015coding} to determine reasons that can be attributed to practitioner's perceptions about the relationship between the development activity metric and defective scripts. As shown in Figure~\ref{fig-meth-rq2-gt}, we first identify `raw text' from the interview. We extract raw text if any portion of the content provided a reason related to practitioner agreement or disagreement for a specific metric. For agreement, we consider agree and strongly agree. For disagreement, we consider disagree and strongly disagree. Next we generate categories from the codes. Finally, we derive the reasons from the identified categories. We report the reasons along with the corresponding metric.


\subsection{Answer to RQ2}
\label{res-rq2} 

We answer RQ2 by providing survey results and results from qualitative analysis in this section. 

\subsubsection{Survey Results}
\label{res-rq2-survey} 

\begin{figure}[]
\centering
\includegraphics[scale=0.45]{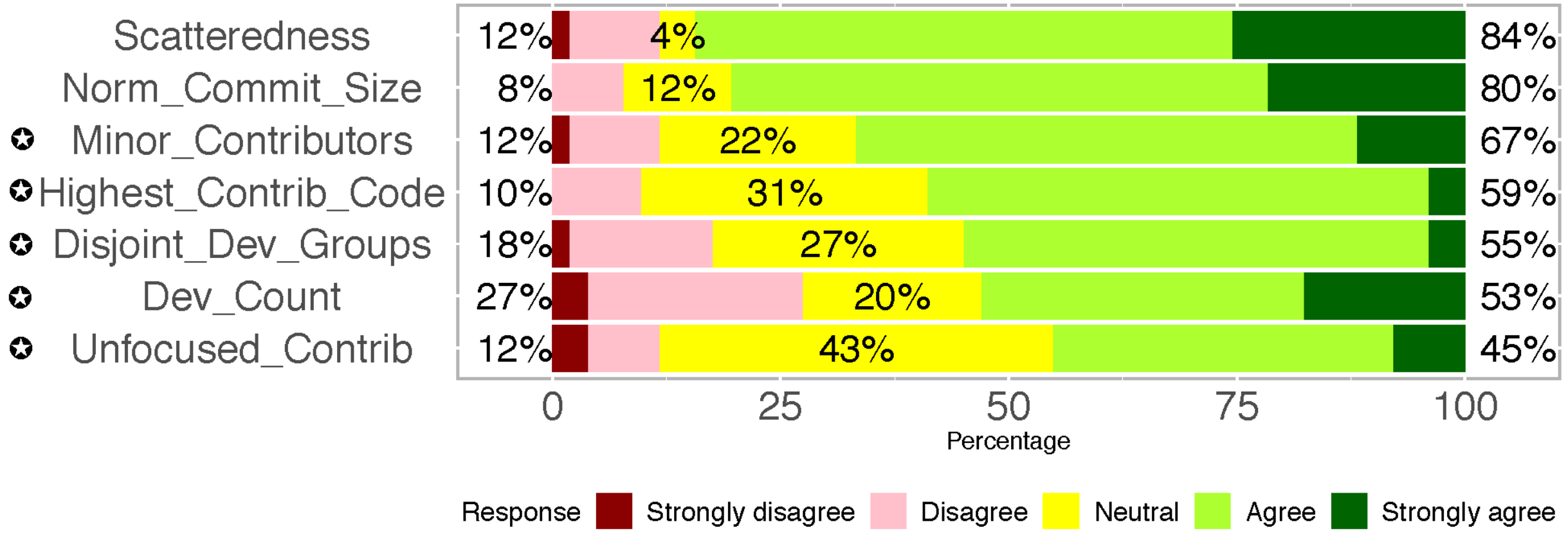}
\caption{Survey findings that illustrate to what extent practitioners agree with the seven metrics and their relationships with defective scripts.} 
\label{fig-survey-res-rq1}
\end{figure}

We obtained a 20.4\% response rate for our survey, which is typical for software engineering-based surveys~\citep{smith:chase2013:survey}. The mean experience in Puppet of the survey respondents is 3.9 years. We present the findings in Figure~\ref{fig-survey-res-rq1}. The percentage values on the right hand side correspond to the percentage of practitioners which responded agree or strongly agree. For example, we observe 84\% of practitioners to agree or strongly agree with scatteredness. At least 80\% of the practitioners who responded, agreed or strongly agreed that scatteredness and normalized commit size are related with quality of IaC scripts. The least agreed upon metric is unfocused contribution with 45\% agreement.       

Of the 51 survey respondents, 11 agreed to participate for semi-structured interviews. Of the 11 interviewees, eight were developers, two were DevOps consultants, and one was a project manager. The mean experience of the 11 interviewees is 5.1 years. The duration of interviews varied from 13 to 29 minutes. Through our qualitative analysis of interview content, we identified a set of reasons that attribute to the agreements or disagreements. For the seven metrics, the first and second author, respectively, obtained 8 and 10 reasons that attribute to practitioner perception. The Cohen's Kappa is 0.87, and we resolve the disagreements for the identified reasons by discussing if additional reasons identified by the second author are similar to that of the first author. Upon resolving disagreements, we obtain eight reasons that are identified both by the first author and second author.   

\subsubsection{Why do practitioners agree and disagree?}
\label{res-rq1-reasons} 

The reasons for agreement and disagreement are respectively underlined using a \ul{solid} and \dotuline{dashed} line. The number of interviewees who \ul{agree} and \dotuline{disagree} are enclosed in a parenthesis. A summary of the reasons on why practitioners agree and disagree is listed in Table~\ref{tab-res-rq2-agr-dis}.

\begin{table}[]
\footnotesize
\captionsetup{justification=centering}
\caption{Summary of Agreement and Disagreement Reasons for the Seven Development Activity Metrics}
\centering
\label{tab-res-rq2-agr-dis}
\begin{tabular}{p{4cm}p{3cm}p{3cm}}
\hline
\textbf{Metric} & \textbf{Agreement} & \textbf{Disagreement} \\
\hline
Developer count &  communication problem & knowledge sharing \\
\hline
Disjointness  between  developer  groups  &  communication problem & knowledge sharing \\
\hline
Highest  contributor's  code  &  lack of context & knowledge sharing \\
\hline
Minor  contributors    &  lack of context & knowledge sharing \\
\hline
Normalized commit size &  code review and debugging difficulty & developer experience \\
\hline
Scatteredness &  code review difficulty & lack of context \\
\hline
Unfocused  contribution  &  distraction & bubble problem \\
\hline
\end{tabular}
\end{table} 

Agreement and disagreement reasons reported by practitioners for the seven identified metrics are stated below:


\textit{Developer count (\ul{8}, \dotuline{3})}: Developer count and defective scripts are related because of \ul{communication problems}, as the complexity of maintaining proper communication may increase. One interviewee stated``\textit{obviously the more people have their hand on the code base, and if good communication isn't happening, then they're gonna have different opinions about how things should end up}''. Brooks~\citep{Brooks:MythicalManMonth:1995} provided a rule of thumb stating``\textit{if there are $n$ workers working on a project...there are potentially almost $2^{n}$ teams within which coordination must occur}'', indicating adding developers to software development will result in increased communication complexity. Disagreeing interviewees stated when developers work on the same script \dotuline{knowledge sharing} happens, which can increase the quality of IaC scripts. One interviewee also reminded about reality:``\textit{I don't think its reasonable to expect that each script is maintained by only a low amount of people. Generally, they [teams] strive to have many people to write as many scripts as possible because that one guy who understands that code can leave}''. 


\textit{Disjointness between developer groups (\ul{7}, \dotuline{3})}: Disjointness between developer groups and defective scripts can be related because of \ul{communication problems} between the groups. According to agreeing interviewees developers have their own opinions, and without coordination scripts are likely to be defective:``\textit{if I need to hop in a Puppet module for my needs, there should be clear communications around the purpose of the module and how updates should occur, I am gonna make updates to serve my purpose but they may have effects outside the module.}''. According to disagreeing practitioners, due to \dotuline{knowledge sharing}, collaboration between developers, even if in disjoint groups, will lead to increased quality of IaC scripts because collaboration leads to knowledge sharing.

\textit{Highest contributor's code (\ul{8}, \dotuline{3})}: Interviewees mentioned \ul{lack of context}: other developers do not have the full the context as the highest contributor, and if the other developers contribute more than the highest contributor, the script suffers from quality issues. On the contrary, due to \dotuline{knowledge sharing}, the highest contributor's code amount may not relate with defective scripts. The highest contributor can share knowledge with other developers on how to modify the script without introducing defects. Such reasoning is consistent with Martin's `collective ownership' strategy~\citep{uncle:bob:clean:coder}, which advocates for any team member to make changes to any software artifact. 

\textit{Minor contributors (\ul{8}, \dotuline{3})}: The relationship between minor contributors and defective scripts can exist because of \ul{lack of context}: developers who contribute less do not have the full context compared with the developers who contribute the majority of the code. Interviewees disagreed for \dotuline{knowledge sharing}, as working on the same script can facilitate sharing of common knowledge needed to modify defects.  

\textit{Normalized commit size (\ul{8}, \dotuline{2})}: Interviewees stated commit size and defective scripts is related for two reasons: \ul{code review difficulty} and \ul{debugging difficulty}. Researchers~\citep{storey:codereview:2018} have reported large-sized commits are harder to review, and susceptible of introducing defects. Their observation are also supported by anecdotal evidence by industry experts in IaC~\citep{kief:iac:book}, who advise in committing small changes at a time. Furthermore, with large-sized commits, bug localization becomes challenging. One interviewee mentioned``\textit{you should always split your changes ... this [large-sized] commit is not helping you to find when and where the bug was introduced; it is harder to fix and revert}''. Reviewing Puppet code can be harder according to one interviewee``\textit{reviewing Puppet code is more challenging than regular code ... if I look for a JavaScript diff I can look side by side and see what the new functions do, whereas with Puppet you have to think it through your head}''. Interviewees also mentioned how large-sized commits are introduced. One interviewee mentioned IaC scripts can be authored by system operators who do not have a software development background:``\textit{Ansible and Puppet modules are written by sysadmins and not developers...we have been in this shift where sysadmins become infrastructure developers and the setup of being an admin to a developer is different because there is a difference between a developer and a sysadmin.}''. Disagreeing interviewees mentioned \dotuline{developer experience}. A developer who is well-versed with the script may be able to submit a large-sized commit without making the script defective. 

\textit{Scatteredness (\ul{7}, \dotuline{3})}: \ul{Code review difficulty} is why interviewees perceive scatteredness is related with defective IaC scripts. If the changes are spread throughout the script, developers may miss a defect while reviewing the code changes. On the contrary, three interviewees disagreed stating \dotuline{lack of context} as the reason. 

\textit{Unfocused contribution (\ul{7}, \dotuline{4})}: \ul{Distraction} is why unfocused contribution is related to defective scripts. Working on multiple scripts can lead to distraction, which could influence the quality of IaC script development. Their reasoning is congruent with Weinberg~\citep{Weinberg:1992:QSM}, who proposed a rule of thumb suggesting a 10\% decrease in software quality whenever developers switch between projects. Interviewees who disagreed stated the \dotuline{bubble problem}--developers may get stuck in a `bubble' if they don't work on multiple scripts. When developers work on multiple scripts, they gather knowledge collected from one script and apply it to another script. One interviewee said ``\textit{working on one script means you are in a bubble, you cannot see what solutions were conceived to solve certain problems}''.    

\textbf{Additional Development Activities Mentioned by Practitioners}: During the interview process the practitioners mentioned additional development activities not included in our set of seven metrics. These activities are: not using version control for IaC scripts, development of IaC scripts without design, undocumented code in IaC scripts, not using feature flags, not using gradual rollouts, inadequate logging, and inadequate testing. 

\textbf{Nuanced perspectives}: At least 80\% of the survey respondents agree that large-sized commits and scatteredness are related to defective IaC scripts. Our quantitative analysis reveals that the relationship of these two metrics with defective scripts are not prevalent, and we refrain ourselves from declaring these two activities as anti-patterns. Practitioner perceptions are often formed by their own experiences, which may not always be supported by actual evidence~\citep{Devanbu:Belief:Evidence}. According to Srikanth and Menzies~\citep{Srikanth:Menzies:TSE2019:DevBeliefs}, ``\textit{documenting developer beliefs should be the start, not the end, of software engineering research. Once prevalent beliefs are found, they should be checked against real-world data}'', suggesting researchers to complement developer perception analysis with software repository data. Our findings suggest a nuanced perspective on developers' perception on anti-patterns, as the relationship between normalized commit size and defective scripts is not as prevalent as practitioners perceive~\citep{kief:iac:book}~\citep{dzone:small:commits}. While considering the use of these development anti-patterns we advise practitioners to consider the nuanced context of perception and quantitative findings.  


\begin{mdframed}
\textbf{Answer to RQ2}: Practitioners show varying agreement on the relationship between seven development activity metrics and defective IaC scripts. They mostly agree with scatteredness and least agree with unfocused contribution.       
\end{mdframed}


\section{RQ3: How can we construct defect prediction models for IaC scripts using development activity metrics?}
\label{rq3} 

Defect prediction models can help practitioners automatically identify scripts that are likely to be defective. Instead of inspecting and testing all IaC scripts used by a team, practitioners from the team can prioritize their inspection and testing efforts for a smaller set of scripts that are likely to be defective. In Sections~\ref{meth-rq3} and~\ref{res-rq3} respectively, we provide the methodology and results for RQ3. 


\subsection{Methodology for RQ3} 
\label{meth-rq3} 

Our procedure to construct models to predict defective scripts can be summarized as following: 

\textit{Metric exclusion}: From statistical analysis we exclude any metric, which shows no relationship with defective IaC scripts for any of the four datasets. Rest of the metrics are used to construct defect prediction models. Our hypothesis that if a metric shows no quantitative relationship with defective scripts for any of the datasets, then metric the may not be useful in constructing defect prediction models. 

\textit{Log transformation}: We first apply log-transformation on the extracted counts for each source code property. Application of log transformation on numerical features help in prediction~\citep{Menzies:TSE2007}.

\textit{Principal Component Analysis (PCA)}: We use principal component analysis (PCA)~\citep{Tan:2005:IDM} to account for multi-collinearity amongst features~\citep{Tan:2005:IDM}. Principal components that account for at least 95\% of the total variance are used as input to statistical learners. 

\textit{Statistical learners}: We use four statistical learners to construct prediction models. These learners are classification and regression trees (CART), logistic regression (LR), Naive Bayes (NB), and Random Forest (RF). CART generates a tree based on the impurity measure, and uses that tree to provide decisions based on input features~\citep{cart:original}. We select CART because this learner does not make any assumption on the distribution of features, and is robust to model overfitting~\citep{Tan:2005:IDM}~\citep{cart:original}. LR estimates the probability that a data point belongs to a certain class, given the values of features~\citep{logit:original}. LR provides good performance for classification if the features are roughly linear~\citep{logit:original}. We select LR because this learner performs well for classification problems~\citep{logit:original} such as defect prediction~\citep{Rahman:2013:ProcessBetter} and fault prediction~\citep{tracy:tse:lrisgood:2012}. The NB classification technique computes the posterior probability of each class to make prediction decisions. We select NB because prior research has reported that defect prediction models that use NB perform well~\citep{tracy:tse:lrisgood:2012}. RF is an ensemble technique that creates multiple classification trees, each of which are generated by taking random subsets of the training data~\citep{Breiman2001:RF:ORIGINAL}~\citep{Tan:2005:IDM}. Unlike LR, RF does not expect features to be linear for good classification performance. Researchers~\citep{Ghotra:ICSE2015} recommended the use of statistical learners that uses ensemble techniques to build defect prediction models. 

\textit{10$\times$10-fold evaluation}: We use 10$\times$10-fold cross validation to evaluate our prediction models. We use the 10$\times$10-fold cross validation evaluation approach by randomly partitioning the dataset into 10 equal sized subsamples or folds~\citep{Tan:2005:IDM}. The performance of the constructed prediction models are tested by using nine of the 10 folds as training data and the remaining fold as test data. We repeat the process of creating training and test data using 10 folds 10 times. 

\textit{Parameter tuning of learners}: Following Fu et al.~\citep{Fu2016} and Tantithamthavorn wt al.~\citep{Chakkrit:ICSE16}'s recommendations we tune the parameters of statistical learners using differential evolution (DE). We select the parameters needed to tune form prior work~\citep{Fu2016}. 

\textit{Comparison with prior approaches}: We compare the development activity-based constructed prediction models with three prior approaches that address quality issues in IaC scripts: (i) the bag-of-word (BOW) approach~\citep{me:icst2018:iac}, (ii) implementation smells~\citep{SharmaPuppet2016} listed in Table~\ref{table:meth:pred:smells}, and (iii) script-level quality metrics~\citep{Bent:Saner2018:Puppet} listed in Table~\ref{table:meth:pred:bent}. 

We use a variant of the \textit{S}cott \textit{K}nott (SK) test~\citep{Chakkrit:TSE2017} to compare prediction performance. This variant of SK does not assume input to be normal, and accounts for negligible effect size~\citep{Chakkrit:TSE2017}. SK uses hierarchical clustering analysis to partition the input data into significantly ($\alpha=0.05$) distinct ranks~\citep{Chakkrit:TSE2017}.

\textit{Performance measures}: We use three performance metrics to evaluate the constructed prediction models: precision, recall, and F-measure. Precision measures the proportion of IaC scripts that are actually defective given that the model predicts as defective. Recall measures the proportion of defective IaC scripts that are correctly predicted by the prediction model. F-measure is the harmonic mean of precision and recall.


\begin{table}[]
\footnotesize
\captionsetup{justification=centering}
\caption{Implementation Smells~\citep{SharmaPuppet2016}}
\centering
\label{table:meth:pred:smells}
\begin{tabular}{p{3cm}p{8cm}}
\hline
\textbf{Smell Name}    & \textbf{Description}  \\
\hline
Missing Default Case   & The default case is missing \\
Inconsistent Naming    & Names deviates from convention recommended by configuration tool vendors \\
Complex Expression     &  The configuration script contains one or many difficult-to-understand complex expressions \\
Duplicate Entity       & The configuration script contains duplicate hash keys or parameters \\
Misplaced Attribute    & Placement of attributes within a resource or a class does not follow a recommended order \\
Improper Alignment     & The code is not properly aligned \\
Invalid Property Value & The configuration script contains invalid value of a property or an attribute \\
Incomplete Tasks       & The configuration scripts include comments that has `fixme' and `todo' as keywords \\
Deprecated Statement   & The configuration script uses one of the deprecated statements \\ 
Improper Quote         & Single and double quotes are misused in the configuration scripts \\
Long Statement         & The configuration script contains long statements\\
Incomplete Conditional & A terminating `else' clause in an if-else block\\
Unguarded Variable     & A variable is not enclosed in braces when being interpolated in a string \\
\hline
\end{tabular}
\end{table} 


\begin{table}[]
\footnotesize
\captionsetup{justification=centering}
\caption{Code Quality Metrics~\citep{Bent:Saner2018:Puppet}}
\centering
\label{table:meth:pred:bent}
\begin{tabular}{p{3cm}p{8cm}}
\hline
\textbf{Quality Metric} & \textbf{Description}  \\
\hline
Filelength    &  Lines in a script \\
Complexity    & Total control statements and  alternatives  in
case statements per script  \\ 
Parameters    & Total parameters per script  \\
Execs         & Total `exec' per script  \\
Lint warnings & Lint warnings per script  \\
Fan-in        & Incoming  dependencies per script \\
\hline
\end{tabular}
\end{table}


\subsection{Answer to RQ3} 
\label{res-rq3}

As shown in Table~\ref{table-res-rq1}, we observe all metrics to show a relation with defective IaC scripts for at least one dataset.  We do not remove any metrics from our set while constructing prediction models. We report the number of principal components used to construct our defect prediction models in Table~\ref{res-rq2-pca}. 

We report the performance of our constructed prediction models to predict defective scripts as following: we compare median precision achieved for 10$\times$10-fold using development activity metrics to that of BOW, code quality, and implementation smells respectively, in Tables~\ref{res:rq1:table:pre:bow},~\ref{res:rq1:table:pre:code}, and~\ref{res:rq1:table:pre:smell}. Tables~\ref{res:rq1:table:rec:bow}, ~\ref{res:rq1:table:rec:code}, and~\ref{res:rq1:table:rec:smell}, respectively, compare the median recall of models using development activity metrics to that with BOW, code quality metrics, and implementation smells. Finally, Tables~\ref{res:rq1:table:f1:bow},~\ref{res:rq1:table:f1:code}, and~\ref{res:rq1:table:f1:smell} respectively compare the median F-measure of models using development activity metrics to that with BOW, code quality metrics, and implementation smells. For all of above the above-mentioned Tables, the shaded cells indicate the highest median precision as determined by SK test.  

With respect to F-measure, development activity metrics are better than the three approaches for three datasets. Using development activity metrics, we construct prediction models that have the highest median F-measure for three datasets. Considering median recall, prediction models with implementation smells provide the best prediction performance for three datasets. With respect to median precision, development activity-based models provide the highest prediction performance for two datasets. Our findings show that prediction models created using development activity metrics are better for predicting which scripts are defective when compared with the BOW, code quality metrics, and implementation smells approaches.      

\begin{table}[]
\centering
\caption{Number of Principal Components used for Prediction Models}
\label{res-rq2-pca}
\footnotesize
{
\begin{tabular}{ p{3.5cm}  p{1cm} p{1cm} p{3.5cm} }
\hline
\textbf{Dataset}   & \textbf{Activity} & \textbf{Code} & \textbf{BOW} \\
\hline
\textbf{Mirantis}  & 4 & 3 &  50 \\
\textbf{Mozilla}   & 4 & 4 &  140\\
\textbf{Openstack} & 4 & 5 &  400\\
\textbf{Wikimedia} & 3 & 4 &  150\\
\hline
\end{tabular}
}
\end{table}


\begin{table*}[]
\centering
\caption{Comparing Median Precision Between Development Activity Metrics and BOW}
\label{res:rq1:table:pre:bow}
\footnotesize{  
\begin{tabular}{p{1.2cm} p{0.9cm} p{0.9cm} p{0.9cm} p{0.9cm} p{0.9cm} p{0.6cm} p{0.6cm} p{0.6cm}  }
\hline
\textbf{Dataset} & \multicolumn{4}{c}{\textbf{Development}} & \multicolumn{4}{c}{\textbf{BOW}}  \\
\hline
 &  \textbf{CART} & \textbf{LR} & \textbf{NB} & \textbf{RF} & \textbf{CART} & \textbf{LR} & \textbf{NB} & \textbf{RF}  \\
\hline
MIR & \cellcolor{lightgray} 0.84 & 0.69 & 0.71 &  0.81                       & 0.62 & 0.63 & 0.75 & 0.72    \\
MOZ & 0.73                       & 0.65 & 0.67 & \cellcolor{lightgray}  0.82 & 0.51 & 0.64 & 0.63 & 0.65    \\
OST & 0.71                       & 0.67 & 0.71 & 0.69                        & 0.63 & 0.58 & \cellcolor{lightgray} 0.74 & 0.57   \\
WIK & 0.71                       & 0.68 & 0.65 & 0.67                        & 0.60 & 0.66 & 0.75 & \cellcolor{lightgray} 0.76   \\
\hline
\end{tabular}
}
\end{table*} 

\begin{table*}[]
\centering
\caption{Comparing Median Precision Between Development Activity Metrics and Code Quality Metrics} 
\label{res:rq1:table:pre:code}
\footnotesize{  
\begin{tabular}{p{1.2cm} p{0.9cm} p{0.9cm} p{0.9cm} p{0.9cm} p{0.9cm} p{0.6cm} p{0.6cm} p{0.6cm}  }
\hline
\textbf{Dataset} & \multicolumn{4}{c}{\textbf{Development}} & \multicolumn{4}{c}{\textbf{Code Quality}}  \\
\hline
 &  \textbf{CART} & \textbf{LR} & \textbf{NB} & \textbf{RF} & \textbf{CART} & \textbf{LR} & \textbf{NB} & \textbf{RF}  \\
\hline
MIR & \cellcolor{lightgray} 0.84 & 0.69 & 0.71 &                        0.81 & 0.75                       & 0.52 & 0.51 & 0.50  \\
MOZ & 0.73                       & 0.65 & 0.67 & \cellcolor{lightgray}  0.82 & 0.74                       & 0.53 & 0.51 & 0.51   \\
OST & 0.71                       & 0.67 & 0.71 & 0.69                        & \cellcolor{lightgray} 0.75 & 0.58 & 0.74 & 0.57   \\
WIK & 0.71                       & 0.68 & 0.65 & 0.67                        & \cellcolor{lightgray} 0.76 & 0.52 & 0.61 & 0.52    \\
\hline
\end{tabular}
}
\end{table*} 

\begin{table*}[]
\centering
\caption{Comparing Median Precision Between Development Activity Metrics and Implementation Smells}
\label{res:rq1:table:pre:smell}
\footnotesize{  
\begin{tabular}{p{1.2cm} p{0.9cm} p{0.9cm} p{0.9cm} p{0.9cm} p{0.9cm} p{0.6cm} p{0.6cm} p{0.6cm}  }
\hline
\textbf{Dataset} & \multicolumn{4}{c}{\textbf{Development}} & \multicolumn{4}{c}{\textbf{Implementation Smell}}  \\
\hline
 &  \textbf{CART} & \textbf{LR} & \textbf{NB} & \textbf{RF} & \textbf{CART} & \textbf{LR} & \textbf{NB} & \textbf{RF}  \\
\hline
MIR & \cellcolor{lightgray} 0.84 & 0.69 & 0.71 &                        0.81 & 0.52 & 0.66 & 0.69 & 0.79   \\
MOZ & 0.73                       & 0.65 & 0.67 & \cellcolor{lightgray}  0.82 & 0.60 & 0.72 & 0.75 & 0.71   \\
OST & 0.71                       & 0.67 & 0.71 & 0.69                        & 0.57 & 0.65 & \cellcolor{lightgray} 0.77 & 0.66   \\
WIK & 0.71                       & 0.68 & 0.65 & 0.67                        & 0.52 & 0.65 & \cellcolor{lightgray} 0.77 & 0.65   \\
\hline
\end{tabular}
}
\end{table*}



\begin{table*}[]
\centering
\caption{Comparing Median Recall between Development Activity Metrics and BOW}
\label{res:rq1:table:rec:bow}
\footnotesize{  
\begin{tabular}{p{1.2cm} p{0.9cm} p{0.9cm} p{0.9cm} p{0.9cm} p{0.9cm} p{0.6cm} p{0.6cm} p{0.6cm}  }
\hline
\textbf{Dataset} & \multicolumn{4}{c}{\textbf{Development}} & \multicolumn{4}{c}{\textbf{BOW}} \\
\hline
 &  \textbf{CART} & \textbf{LR} & \textbf{NB} & \textbf{RF} & \textbf{CART} & \textbf{LR} & \textbf{NB} & \textbf{RF} \\
\hline
MIR & \cellcolor{lightgray} 0.89 & 0.86 & 0.86 & 0.88                       & 0.69 & 0.53 & 0.67 & 0.84   \\
MOZ & \cellcolor{lightgray} 0.70 & 0.55 & 0.42 & \cellcolor{lightgray} 0.70 & 0.25 & 0.41 & 0.51 & 0.65   \\
OST & 0.88 & 0.76 & 0.60 & \cellcolor{lightgray} 0.90                       & 0.62 & 0.61 & 0.71 & 0.70    \\
WIK & 0.68 & 0.59 & 0.56 & 0.76                                             & 0.65 & 0.47 & 0.75 & \cellcolor{lightgray} 0.80   \\
\hline
\end{tabular}
}
\end{table*}  

\begin{table*}[]
\centering
\caption{Comparing Median Recall between Development Activity Metrics and Code Quality Metrics}
\label{res:rq1:table:rec:code}
\footnotesize{  
\begin{tabular}{p{1.2cm} p{0.9cm} p{0.9cm} p{0.9cm} p{0.9cm} p{0.9cm} p{0.6cm} p{0.6cm} p{0.6cm}  }
\hline
\textbf{Dataset} & \multicolumn{4}{c}{\textbf{Development}} & \multicolumn{4}{c}{\textbf{Code Quality}} \\
\hline
 &  \textbf{CART} & \textbf{LR} & \textbf{NB} & \textbf{RF} & \textbf{CART} & \textbf{LR} & \textbf{NB} & \textbf{RF} \\
\hline
MIR & \cellcolor{lightgray} 0.89 & 0.86 & 0.86 & 0.88                       & 0.78 & 0.71 & 0.86 & 0.77   \\
MOZ & \cellcolor{lightgray} 0.70 & 0.55 & 0.42 & \cellcolor{lightgray} 0.70 & 0.69 & 0.59 & 0.42 & 0.67   \\
OST & 0.88                       & 0.76 & 0.60 & 0.90                       & 0.84 & 0.73 & 0.60 & \cellcolor{lightgray} 0.96   \\
WIK & 0.68                       & 0.59 & 0.56 & \cellcolor{lightgray}0.76  & 0.75 & 0.74 & 0.56 & 0.75   \\
\hline
\end{tabular}
}
\end{table*}  

\begin{table*}[]
\centering
\caption{Comparing Median Recall between Development Activity Metrics and Implementation Smells} 
\label{res:rq1:table:rec:smell}
\footnotesize{  
\begin{tabular}{p{1.2cm} p{0.9cm} p{0.9cm} p{0.9cm} p{0.9cm} p{0.9cm} p{0.6cm} p{0.6cm} p{0.6cm}  }
\hline
\textbf{Dataset} & \multicolumn{4}{c}{\textbf{Development}} & \multicolumn{4}{c}{\textbf{Implementation Smell}} \\
\hline
 &  \textbf{CART} & \textbf{LR} & \textbf{NB} & \textbf{RF} & \textbf{CART} & \textbf{LR} & \textbf{NB} & \textbf{RF} \\
\hline
MIR & 0.89                       & 0.86 & 0.86 & 0.88                        & 0.90 &  \cellcolor{lightgray} 0.96 & 0.23 & 0.86   \\
MOZ & \cellcolor{lightgray} 0.70 & 0.55 & 0.42 & \cellcolor{lightgray} 0.70  & 0.47 & 0.41 & 0.32 & 0.49    \\
OST & 0.88                       & 0.76 & 0.60 & 0.90                        & 0.84 & \cellcolor{lightgray} 0.99 & 0.09 & 0.96    \\
WIK & 0.68                       & 0.59 & 0.56 & 0.76                        & \cellcolor{lightgray} 0.93 &  0.90 & 0.05 & 0.81   \\
\hline
\end{tabular}
}
\end{table*}



\begin{table*}[]
\centering
\caption{Comparing Median F-measure between Development Activity Metrics and BOW}
\label{res:rq1:table:f1:bow}
\footnotesize{  
\begin{tabular}{p{1.2cm} p{0.9cm} p{0.9cm} p{0.9cm} p{0.9cm} p{0.9cm} p{0.6cm} p{0.6cm} p{0.6cm}  }
\hline
\textbf{Dataset} & \multicolumn{4}{c}{\textbf{Development}} & \multicolumn{4}{c}{\textbf{BOW}}  \\
\hline
 &  \textbf{CART} & \textbf{LR} & \textbf{NB} & \textbf{RF} & \textbf{CART} & \textbf{LR} & \textbf{NB} & \textbf{RF} \\
\hline
MIR & \cellcolor{lightgray} 0.80 & 0.77 & 0.78 & \cellcolor{lightgray} 0.80 & 0.66 & 0.59 & 0.67 & 0.77    \\
MOZ & 0.68                       & 0.60 & 0.52 & \cellcolor{lightgray} 0.71 & 0.34 & 0.49 & 0.49 & 0.64   \\
OST & \cellcolor{lightgray} 0.74 & 0.71 & 0.66 & \cellcolor{lightgray} 0.74 & 0.62 & 0.65 & 0.69 & 0.67  \\
WIK & 0.67                       & 0.63 & 0.61 & 0.68                       & 0.63 & 0.56 & 0.71 & \cellcolor{lightgray} 0.72  \\
\hline
\end{tabular}
}
\end{table*}  

\begin{table*}[]
\centering
\caption{Comparing median F-measure between Development Activity Metrics and Code Quality Metrics}
\label{res:rq1:table:f1:code}
\footnotesize{  
\begin{tabular}{p{1.2cm} p{0.9cm} p{0.9cm} p{0.9cm} p{0.9cm} p{0.9cm} p{0.6cm} p{0.6cm} p{0.6cm}  }
\hline
\textbf{Dataset} & \multicolumn{4}{c}{\textbf{Development}} & \multicolumn{4}{c}{\textbf{Code quality}}  \\
\hline
 &  \textbf{CART} & \textbf{LR} & \textbf{NB} & \textbf{RF} & \textbf{CART} & \textbf{LR} & \textbf{NB} & \textbf{RF} \\
\hline
MIR & \cellcolor{lightgray} 0.80 & 0.77 & 0.78 & \cellcolor{lightgray} 0.80 & 0.73 & 0.68 & 0.71 & 0.75   \\
MOZ & 0.68                       & 0.60 & 0.52 & \cellcolor{lightgray} 0.71 & 0.68 & 0.65 & 0.63 & 0.66   \\
OST & \cellcolor{lightgray} 0.74 & 0.71 & 0.66 & \cellcolor{lightgray} 0.74 & 0.70 & 0.69 & 0.59 & 0.72   \\
WIK & 0.67 & 0.63 & 0.61 & 0.68                                             & 0.72 & 0.72 & 0.63 & \cellcolor{lightgray}  0.73   \\
\hline
\end{tabular}
}
\end{table*}

\begin{table*}[]
\centering
\caption{Comparing Median F-measure between Development Activity Metrics and Implementation Smells}
\label{res:rq1:table:f1:smell}
\footnotesize{  
\begin{tabular}{p{1.2cm} p{0.9cm} p{0.9cm} p{0.9cm} p{0.9cm} p{0.9cm} p{0.6cm} p{0.6cm} p{0.6cm}  }
\hline
\textbf{Dataset} & \multicolumn{4}{c}{\textbf{Development}} & \multicolumn{4}{c}{\textbf{Implementation Smell}}  \\
\hline
 &  \textbf{CART} & \textbf{LR} & \textbf{NB} & \textbf{RF} & \textbf{CART} & \textbf{LR} & \textbf{NB} & \textbf{RF} \\
\hline
MIR & \cellcolor{lightgray} 0.80 & 0.77 & 0.78 & \cellcolor{lightgray} 0.80 & 0.66 & 0.68 & 0.31 & 0.63   \\
MOZ & 0.68                       & 0.60 & 0.52 & \cellcolor{lightgray} 0.71 & 0.53 & 0.53 & 0.45 & 0.53   \\
OST & \cellcolor{lightgray} 0.74 & 0.71 & 0.66 & \cellcolor{lightgray} 0.74 & 0.71 & 0.73 & 0.16 & 0.72   \\
WIK & 0.67                       & 0.63 & 0.61 & \cellcolor{lightgray} 0.68 & 0.67 & 0.66 & 0.09 & 0.63  \\
\hline
\end{tabular}
}
\end{table*}  

\textbf{Implications of Prediction Models}: With respect to money, personnel, and time, inspection and testing of software source files is an expensive procedure, which should be allocated efficiently~\citep{ostrand:issta2004}~\citep{Adams:PrioritizeUnitTests:2011}~\citep{assmann:2012:test:priority}. Practitioners who maintain and develop IaC scripts need to prioritize what IaC scripts need more inspection and testing. Our constructed prediction models lay the groundwork for (i) providing practitioners the opportunity to automatically identify which IaC scripts are likely to have a defect, and (ii) future research that can build on top of our work to quantify usability of IaC defect prediction models. Furthermore, the model is constructed with development activity metrics, which indicates that regardless of the language the model can be used to predict which scripts are likely to be defective. 




\begin{mdframed}
\textbf{Answer to RQ3}: With respect to F-measure, defect prediction models constructed with development activity metrics outperform three approaches: bag of words (BOW), implementation smells, and code quality metrics. 
\end{mdframed}


\section{Discussion}
\label{discussion}

In this section, we discuss our findings by first synthesizing the development activities that relate with defective scripts into development anti-patterns. We also discuss future work in this section. 

\paragraph{Development Anti-patterns}
\label{only-anti-pattern}

We identify a set of development anti-patterns based on two criteria: each activity must be (i) related to software quality as reported in prior literature and (ii) supported by quantitative evidence across all four datasets (Section~\ref{res-rq1}). We alphabetically list the identified five development anti-patterns with definitions, prior literature that supports an activity, and possible solutions. The presented possible solutions are recommendations from authors that are subject to empirical validation.  

\begin{shaded}

\hspace{-0.65cm} \textbf{Anti-pattern\#1: Boss is Not Around} \\
\textit{Definition}: The highest contributor does not author all lines of an IaC script. \\ 
\textit{Observation}: The highest contributor may have the full context of the script, which other developers may not have. If the other developers contribute more than the highest contributor, the corresponding script may be defective. On average, the highest contributor contributes 80\%$\mathtt{\sim}$90\% of the code for neutral scripts.   \\
\textit{Prior literature}: Bird et al.~\citep{Bird:Minor}, Rahman and Devanbu~\citep{Rahman:2013:ProcessBetter} \\
\textit{Metric}: Highest contributor's code ($HIGHEST\_CONTRIB\_CODE$)\\
\textit{Solution}: Perform team planning so that the highest contributor can modify and/or verify all the content added to a script. \\

\hspace{-0.65cm}\textbf{Anti-pattern\#2: Many Cooks Spoil}  \\
\textit{Definition}: Having multiple developers working on the same script.\\ 
\textit{Observation}: Defective scripts are modified by 12$\mathtt{\sim}$43 developers, whereas, neutral scripts modified by no more than 11 developers. \\
\textit{Prior literature}: Meneely and Williams~\citep{Meneely:Linus}, Businge et al.~\citep{Foutse:Devs:Android}  \\
\textit{Metric}: Developer count \\
\textit{Solution}: Thoroughly inspect scripts modified by multiple developers. \\

\hspace{-0.65cm}\textbf{Anti-pattern\#3: Minors are Spoilers} \\
\textit{Definition}: The activity of a script being modified by developer(s) who writes no more than 5\% code of the script. \\ 
\textit{Observation}: A neutral script may be modified by at most seven minor contributors, whereas, defective scripts can be modified by 8$\mathtt{\sim}$36 minor contributors. \\
\textit{Prior literature}: Bird et al.~\citep{Bird:Minor}, Rahman and Devanbu~\citep{Rahman:2013:ProcessBetter} \\
\textit{Metric}: Minor contributors \\
\textit{Solution}: Thoroughly inspect scripts modified by multiple minor contributors.  \\

\hspace{-0.65cm}\textbf{Anti-pattern\#4: Silos} \\
\textit{Definition}: The activity of developers working in disjoint groups. \\ 
\textit{Observation}:  Defective scripts are modified by developer groups, which are 1.3$\mathtt{\sim}$2.0 times more disjoint, on average, compared to that of neutral scripts. Even though practitioners perceive IaC as a tool to break silos~\citep{silo:iac:devops}, our quantitative findings suggest that silos exist in IaC development. \\ 
\textit{Prior literature}: Meneely and Williams~\citep{Meneely:Linus}  \\
\textit{Metric}: Disjointness in developer groups \\
\textit{Solution}: Collaborate more instead of working in disjoint groups in IaC development. \\

\hspace{-0.6cm} \textbf{Anti-pattern\#5: Unfocused Contribution} \\
\textit{Definition}: The activity of developers working on an IaC script, who also modify other scripts. \\ 
\textit{Observation}:  On average, for defective scripts unfocused contribution is 62.5\%$\mathtt{\sim}$95.4\% higher than neutral scripts.   \\ 
\textit{Prior literature}: Meneely and Williams~\citep{Meneely:Linus}  \\
\textit{Metric}: Unfocused contribution \\
\textit{Solution}: Perform team planning so that developers can focus on developing one script within a certain time period, instead of multiple developers working simultaneously. \\
\end{shaded} 

\paragraph{Future Work}: We have identified a list of reasons that attribute to practitioner perception in Section~\ref{res-rq2}. Future researchers can investigate to what extent these findings generalize by conducting large-scale quantitative studies. Furthermore, researchers can investigate the synergies between the development activities within themselves.

We have provided a list of possible solutions that are based on author judgement, and not validated by empirical research. Researchers can investigate if our suggested solutions can be validated empirically.   

Practitioners have listed a set of development anti-patterns that are not included in our set of five: not using version control for IaC scripts, development of IaC scripts without design, undocumented code in IaC scripts, not using feature flags, not using gradual rollouts, inadequate logging, and inadequate testing. We urge researchers to investigate if practitioner-mentioned anti-patterns are substantiated by empirical evidence. Such investigation can also yield empirical studies that can characterize other development aspects of IaC, such as usage frequency of logging, testing, and version control. For example, researchers can investigate if testing anti-patterns~\citep{GAROUSI:TEST:SMELLS} and logging anti-patterns~\citep{chen:logging:smells} that are applicable for general purpose programming languages, also apply for IaC.  

Our prediction performance results from Section~\ref{res-rq3} suggest that one approach i.e., development activity metrics or BOW or code quality metrics is not comprehensive because depending on prediction performance metric, one single approach does not provide the best results. Future researchers can investigate what techniques are needed to obtain prediction models that perform better. Furthermore, researchers can investigate how to build and evaluate prediction models that also includes cost-effective measures as used in defect prediction research~\citep{lionel:jss:dp}.


\section{Threats to Validity}
\label{threats}

We discuss the limitations of our paper as following:

\textbf{Conclusion Validity}: Our findings are dependent on the four datasets, which are constructed by raters. The data construction process is susceptible to human judgment, and the raters' experience can bias the identified defective IaC scripts. The process of determining the reasons for practitioner perception are susceptible to rater bias. In both cases, we mitigate these threats by assigning at least two raters. 

Our selection thresholds can be limiting. For example, a repository may contain sufficient amount of Ansible or Chef scripts, but maintained by one practitioner. Such repositories even though active, will be excluded in our analysis based on criteria mention in Section~\ref{dataset}. 

When developing IaC scripts, other socio-technical factors may contribute to the quality of IaC scripts that are not captured by our statistical analysis and metrics. We mitigate this limitation by conducting a univariate and a multivariate analysis to establish the correlation between the seven development activity metrics and defective scripts. 

\textbf{Construct validity}: Use of raters to identify defect-related commits is susceptible to mono-method bias, where subjective judgment of raters can influence the findings. We mitigated this threat by using at least two raters and a resolver.  

For two datasets, Mirantis and Wikimedia, we used raters who are graduate students who determined defect-related commits as part of their class work. Students who participated in the process can be subject to evaluation apprehension, i.e. consciously or sub-consciously relating their performance with the grades they would achieve for the course. We mitigated this threat by clearly explaining to the students that their rating would not affect their grades. 

The raters involved in the categorization process had professional experience in software engineering for at two years on average. Their experience in software engineering may make the raters curious about the expected outcomes of the categorization process, which may effect the distribution of the categorization process. Furthermore, the resolver also has professional experience in software engineering and IaC script development, which could influence the outcome of the defect category distribution.

\textbf{External Validity}: Our findings are subject to external validity, as our findings may not be generalizable for the proprietary domain. The identified reasons and development anti-patterns also may not be comprehensive. Furthermore, our datasets include Puppet scripts, and our derived anti-patterns may not generalize to Ansible or Chef scripts. Also, our sample size of survey respondents is not comprehensive and may not be representative of all IaC practitioners. 

\textbf{Internal Validity}: We have used a combination of commit messages and issue report descriptions to determine if an IaC script is defective. We acknowledge that these messages might not have given the full context for the raters. Our set of metrics is also not comprehensive.


\section{Conclusion}
\label{conclusion}

Defects in IaC scripts can cause serious consequences e.g. creating wide-scale outages. Through systematic investigation, we can identify development anti-patterns for IaC scripts that can be used to advise practitioners on how to improve the quality of IaC scripts. We apply quantitative analysis on 2,138 IaC scripts to determine development anti-patterns. We identify five development anti-patterns, namely, `boss is not around', `many cooks spoil', `minors are spoilers', `silos', and `unfocused contribution'. Our findings show defective scripts are related to development activities considered as software engineering best practices, such as the number of developers working on a script. Our identified development anti-patterns suggest the importance of `as code' activities i.e., application of recommended software development activities for IaC script, as inadequate application of recommended development activities are empirically related to defective IaC scripts. We hope our paper will facilitate further research in the domain of IaC script quality.         



\begin{acknowledgements}
The NSA Science of Security Lablet (award H98230-17-D-0080) at the North Carolina State University supported this research study. We thank the Realsearch research group members for their useful feedback. We also thank the practitioners who answered our questions.
\end{acknowledgements}


\balance 
\bibliographystyle{spbasic} 
\bibliography{process}      

\end{document}